\documentclass[VANCOUVER,STIX1COL]{WileyNJD-v2}
\articletype{Article Type}%

\usepackage{ifxetex}
\ifxetex
\else
    \usepackage[T1]{fontenc}
    \usepackage[utf8]{inputenc}
\fi

\usepackage{hyperref}
\hypersetup{
	pdftitle={Quality tetrahedral mesh generation with HXT and the Growing SPR operation},  
	pdfauthor={Célestin Marot, Jean-François Remacle}, 
	pdfsubject={Mesh Generation},                  
	pdfcreator={Célestin Marot},                   
	pdfkeywords=                                   
	{
		parallel,
		tetrahedral,
		mesh,
		improvement
	},
}

\usepackage{subcaption}
\usepackage{tikz}
\usepackage{pgfplots}
\pgfplotsset{compat=newest}
\pgfplotsset{
    discard if not/.style 2 args={
        x filter/.code={
            \edef\tempa{\thisrow{#1}}
            \edef\tempb{#2}
            \ifx\tempa\tempb
            \else
                
            \fi
        }
    }
}

\definecolor{tolBlack}{HTML}{000000}
\definecolor{tolWhite}{HTML}{FFFFFF}

\definecolor{tolLine0}{HTML}{4477AA}
\definecolor{tolLine1}{HTML}{66CCEE}
\definecolor{tolLine2}{HTML}{228833}
\definecolor{tolLine3}{HTML}{CCBB44}
\definecolor{tolLine4}{HTML}{EE6677}
\definecolor{tolLine5}{HTML}{AA3377}
\definecolor{tolLine6}{HTML}{BBBBBB}

\definecolor{tolHC0}{HTML}{FFFFFF}
\definecolor{tolHC1}{HTML}{DDAA33}
\definecolor{tolHC2}{HTML}{BB5566}
\definecolor{tolHC3}{HTML}{004488}
\definecolor{tolHC4}{HTML}{000000}

\definecolor{tolBlue}{HTML}{0077BB}
\definecolor{tolCyan}{HTML}{33BBEE}
\definecolor{tolTeal}{HTML}{009988}
\definecolor{tolOrange}{HTML}{EE7733}
\definecolor{tolRed}{HTML}{CC3311}
\definecolor{tolMagenta}{HTML}{EE3377}
\definecolor{tolGrey}{HTML}{BBBBBB}

\definecolor{tolMutedIndigo}{HTML}{332288}
\definecolor{tolMutedCyan}{HTML}{88CCEE}
\definecolor{tolMutedTeal}{HTML}{44AA99}
\definecolor{tolMutedGreen}{HTML}{117733}
\definecolor{tolMutedOlive}{HTML}{999933}
\definecolor{tolMutedSand}{HTML}{DDCC77}
\definecolor{tolMutedRose}{HTML}{CC6677}
\definecolor{tolMutedWine}{HTML}{882255}
\definecolor{tolMutedPurple}{HTML}{AA4499}
\definecolor{tolPaleGrey}{HTML}{DDDDDD}

\definecolor{tolPaleBlue}{HTML}{BBCCEE}
\definecolor{tolPaleCyan}{HTML}{CCEEFF}
\definecolor{tolPaleGreen}{HTML}{CCDDAA}
\definecolor{tolPaleYellow}{HTML}{EEEEBB}
\definecolor{tolPaleRed}{HTML}{FFCCCC}
\definecolor{tolDarkBlue}{HTML}{222255}
\definecolor{tolDarkCyan}{HTML}{225555}
\definecolor{tolDarkGreen}{HTML}{225522}
\definecolor{tolDarkYellow}{HTML}{666633}
\definecolor{tolDarkRed}{HTML}{663333}
\definecolor{tolDarkGrey}{HTML}{555555}

\definecolor{tolLightBlue}{HTML}{77AADD}
\definecolor{tolLightCyan}{HTML}{99DDFF}
\definecolor{tolMint}{HTML}{44BB99}
\definecolor{tolPear}{HTML}{BBCC33}
\definecolor{tolOlive}{HTML}{AAAA00}
\definecolor{tolLightYellow}{HTML}{EEDD88}
\definecolor{tolLightOrange}{HTML}{EE8866}
\definecolor{tolPink}{HTML}{FFAABB}

\definecolor{tolSunset0}{HTML}{364B9A}
\definecolor{tolSunset1}{HTML}{4A7BB7}
\definecolor{tolSunset2}{HTML}{6EA6CD}
\definecolor{tolSunset3}{HTML}{98CAE1}
\definecolor{tolSunset4}{HTML}{C2E4EF}
\definecolor{tolSunset5}{HTML}{EAECCC}
\definecolor{tolSunset6}{HTML}{FEDA8B}
\definecolor{tolSunset7}{HTML}{FDB366}
\definecolor{tolSunset8}{HTML}{F67E4B}
\definecolor{tolSunset9}{HTML}{DD3D2D}
\definecolor{tolSunset10}{HTML}{A50026}
\definecolor{tolSunsetError}{HTML}{FFFFFF}

\definecolor{tolBuRd0}{HTML}{2166AC}
\definecolor{tolBuRd1}{HTML}{4393C3}
\definecolor{tolBuRd2}{HTML}{92C5DE}
\definecolor{tolBuRd3}{HTML}{D1E5F0}
\definecolor{tolBuRd4}{HTML}{F7F7F7}
\definecolor{tolBuRd5}{HTML}{FDDBC7}
\definecolor{tolBuRd6}{HTML}{F4A582}
\definecolor{tolBuRd7}{HTML}{D6604D}
\definecolor{tolBuRd8}{HTML}{B2182B}
\definecolor{tolBuRdError}{HTML}{FFEE99}

\definecolor{tolPRGn0}{HTML}{762A83}
\definecolor{tolPRGn1}{HTML}{9970AB}
\definecolor{tolPRGn2}{HTML}{C2A5CF}
\definecolor{tolPRGn3}{HTML}{E7D4E8}
\definecolor{tolPRGn4}{HTML}{F7F7F7}
\definecolor{tolPRGn5}{HTML}{D9F0D3}
\definecolor{tolPRGn6}{HTML}{ACD39E}
\definecolor{tolPRGn7}{HTML}{5AAE61}
\definecolor{tolPRGn8}{HTML}{1B7837}
\definecolor{tolPRGnError}{HTML}{FFEE99}

\definecolor{tolYlOrBr0}{HTML}{FFFFE5}
\definecolor{tolYlOrBr1}{HTML}{FFF7BC}
\definecolor{tolYlOrBr2}{HTML}{FEE391}
\definecolor{tolYlOrBr3}{HTML}{FEC44F}
\definecolor{tolYlOrBr4}{HTML}{FB9A29}
\definecolor{tolYlOrBr5}{HTML}{EC7014}
\definecolor{tolYlOrBr6}{HTML}{CC4C02}
\definecolor{tolYlOrBr7}{HTML}{993404}
\definecolor{tolYlOrBr8}{HTML}{662506}
\definecolor{tolYlOrBrError}{HTML}{888888}

\definecolor{tolIridescent0}{HTML}{FEFBE9}
\definecolor{tolIridescent1}{HTML}{FCF7D5}
\definecolor{tolIridescent2}{HTML}{F5F3C1}
\definecolor{tolIridescent3}{HTML}{EAF0B5}
\definecolor{tolIridescent4}{HTML}{DDECBF}
\definecolor{tolIridescent5}{HTML}{D0E7CA}
\definecolor{tolIridescent6}{HTML}{C2E3D2}
\definecolor{tolIridescent7}{HTML}{B5DDD8}
\definecolor{tolIridescent8}{HTML}{A8D8DC}
\definecolor{tolIridescent9}{HTML}{9BD2E1}
\definecolor{tolIridescent10}{HTML}{8DCBE4}
\definecolor{tolIridescent11}{HTML}{81C4E7}
\definecolor{tolIridescent12}{HTML}{7BBCE7}
\definecolor{tolIridescent13}{HTML}{7EB2E4}
\definecolor{tolIridescent14}{HTML}{88A5DD}
\definecolor{tolIridescent15}{HTML}{9398D2}
\definecolor{tolIridescent16}{HTML}{9B8AC4}
\definecolor{tolIridescent17}{HTML}{9D7DB2}
\definecolor{tolIridescent18}{HTML}{9A709E}
\definecolor{tolIridescent19}{HTML}{906388}
\definecolor{tolIridescent20}{HTML}{805770}
\definecolor{tolIridescent21}{HTML}{684957}
\definecolor{tolIridescent22}{HTML}{46353A}
\definecolor{tolIridescentError}{HTML}{999999}

\definecolor{tolRainbowWhBr0}{HTML}{E8ECFB}
\definecolor{tolRainbowWhBr1}{HTML}{DDD8EF}
\definecolor{tolRainbowWhBr2}{HTML}{D1C1E1}
\definecolor{tolRainbowWhBr3}{HTML}{C3A8D1}
\definecolor{tolRainbowWhBr4}{HTML}{B58FC2}
\definecolor{tolRainbowWhBr5}{HTML}{A778B4}
\definecolor{tolRainbowWhBr6}{HTML}{9B62A7}
\definecolor{tolRainbowWhBr7}{HTML}{8C4E99}
\definecolor{tolRainbowWhBr8}{HTML}{6F4C9B}
\definecolor{tolRainbowWhBr9}{HTML}{6059A9}
\definecolor{tolRainbowWhBr10}{HTML}{5568B8}
\definecolor{tolRainbowWhBr11}{HTML}{4E79C5}
\definecolor{tolRainbowWhBr12}{HTML}{4D8AC6}
\definecolor{tolRainbowWhBr13}{HTML}{4E96BC}
\definecolor{tolRainbowWhBr14}{HTML}{549EB3}
\definecolor{tolRainbowWhBr15}{HTML}{59A5A9}
\definecolor{tolRainbowWhBr16}{HTML}{60AB9E}
\definecolor{tolRainbowWhBr17}{HTML}{69B190}
\definecolor{tolRainbowWhBr18}{HTML}{77B77D}
\definecolor{tolRainbowWhBr19}{HTML}{8CBC68}
\definecolor{tolRainbowWhBr20}{HTML}{A6BE54}
\definecolor{tolRainbowWhBr21}{HTML}{BEBC48}
\definecolor{tolRainbowWhBr22}{HTML}{D1B541}
\definecolor{tolRainbowWhBr23}{HTML}{DDAA3C}
\definecolor{tolRainbowWhBr24}{HTML}{E49C39}
\definecolor{tolRainbowWhBr25}{HTML}{E78C35}
\definecolor{tolRainbowWhBr26}{HTML}{E67932}
\definecolor{tolRainbowWhBr27}{HTML}{E4632D}
\definecolor{tolRainbowWhBr28}{HTML}{DF4828}
\definecolor{tolRainbowWhBr29}{HTML}{DA2222}
\definecolor{tolRainbowWhBr30}{HTML}{B8221E}
\definecolor{tolRainbowWhBr31}{HTML}{95211B}
\definecolor{tolRainbowWhBr32}{HTML}{721E17}
\definecolor{tolRainbowWhBr33}{HTML}{521A13}
\definecolor{tolRainbowWhBrError}{HTML}{666666}

\definecolor{tolRainbowDiscrete0}{HTML}{E8ECFB}
\definecolor{tolRainbowDiscrete1}{HTML}{D9CCE3}
\definecolor{tolRainbowDiscrete2}{HTML}{D1BBD7}
\definecolor{tolRainbowDiscrete3}{HTML}{CAACCB}
\definecolor{tolRainbowDiscrete4}{HTML}{BA8DB4}
\definecolor{tolRainbowDiscrete5}{HTML}{AE76A3}
\definecolor{tolRainbowDiscrete6}{HTML}{AA6F9E}
\definecolor{tolRainbowDiscrete7}{HTML}{994F88}
\definecolor{tolRainbowDiscrete8}{HTML}{882E72}
\definecolor{tolRainbowDiscrete9}{HTML}{1965B0}
\definecolor{tolRainbowDiscrete10}{HTML}{437DBF}
\definecolor{tolRainbowDiscrete11}{HTML}{5289C7}
\definecolor{tolRainbowDiscrete12}{HTML}{6195CF}
\definecolor{tolRainbowDiscrete13}{HTML}{7BAFDE}
\definecolor{tolRainbowDiscrete14}{HTML}{4EB265}
\definecolor{tolRainbowDiscrete15}{HTML}{90C987}
\definecolor{tolRainbowDiscrete16}{HTML}{CAE0AB}
\definecolor{tolRainbowDiscrete17}{HTML}{F7F056}
\definecolor{tolRainbowDiscrete18}{HTML}{F7CB45}
\definecolor{tolRainbowDiscrete19}{HTML}{F6C141}
\definecolor{tolRainbowDiscrete20}{HTML}{F4A736}
\definecolor{tolRainbowDiscrete21}{HTML}{F1932D}
\definecolor{tolRainbowDiscrete22}{HTML}{EE8026}
\definecolor{tolRainbowDiscrete23}{HTML}{E8601C}
\definecolor{tolRainbowDiscrete24}{HTML}{E65518}
\definecolor{tolRainbowDiscrete25}{HTML}{DC050C}
\definecolor{tolRainbowDiscrete26}{HTML}{A5170E}
\definecolor{tolRainbowDiscrete27}{HTML}{72190E}
\definecolor{tolRainbowDiscrete28}{HTML}{42150A}
\definecolor{tolRainbowDiscreteError}{HTML}{777777}

\pgfplotsset{%
	/pgfplots/@@brewer set cycle list/.code={%
		\pgfkeysalso{/pgfplots/ensure colormap={/pgfplots/colormap/#1},/pgfplots/cycle list name=#1}%
	},
	/pgfplots/colormap/tolLines/.style={
		colormap={tolLines}{
			color=(tolLine0);
			color=(tolLine1);
			color=(tolLine2);
			color=(tolLine3);
			color=(tolLine4);
			color=(tolLine5);
			color=(tolLine6);
		},
		cycle list/.define={tolLines}{[of colormap=tolLines]},
	},
	/pgfplots/cycle list/tolLines/.style={/pgfplots/@@brewer set cycle list={tolLines}},
	/pgfplots/colormap/tolHC/.style={
		colormap={tolHC}{
			color=(tolHC0);
			color=(tolHC1);
			color=(tolHC2);
			color=(tolHC3);
			color=(tolHC4);
		},
		cycle list/.define={tolHC}{[of colormap=tolHC]},
	},
	/pgfplots/cycle list/tolHC/.style={/pgfplots/@@brewer set cycle list={tolHC}},
	/pgfplots/colormap/tolSunset/.style={
		colormap={tolSunset}{
			color=(tolSunset0);
			color=(tolSunset1);
			color=(tolSunset2);
			color=(tolSunset3);
			color=(tolSunset4);
			color=(tolSunset5);
			color=(tolSunset6);
			color=(tolSunset7);
			color=(tolSunset8);
			color=(tolSunset9);
			color=(tolSunset10);
		},
		cycle list/.define={tolSunset}{[of colormap=tolSunset]},
	},
	/pgfplots/cycle list/tolSunset/.style={/pgfplots/@@brewer set cycle list={tolSunset}},
	/pgfplots/colormap/tolBuRd/.style={
		colormap={tolBuRd}{
			color=(tolBuRd0);
			color=(tolBuRd1);
			color=(tolBuRd2);
			color=(tolBuRd3);
			color=(tolBuRd4);
			color=(tolBuRd5);
			color=(tolBuRd6);
			color=(tolBuRd7);
			color=(tolBuRd8);
		},
		cycle list/.define={tolBuRd}{[of colormap=tolBuRd]},
	},
	/pgfplots/cycle list/tolBuRd/.style={/pgfplots/@@brewer set cycle list={tolBuRd}},
	/pgfplots/colormap/tolPRGn/.style={
		colormap={tolPRGn}{
			color=(tolPRGn0);
			color=(tolPRGn1);
			color=(tolPRGn2);
			color=(tolPRGn3);
			color=(tolPRGn4);
			color=(tolPRGn5);
			color=(tolPRGn6);
			color=(tolPRGn7);
			color=(tolPRGn8);
		},
		cycle list/.define={tolPRGn}{[of colormap=tolPRGn]},
	},
	/pgfplots/cycle list/tolPRGn/.style={/pgfplots/@@brewer set cycle list={tolPRGn}},
	/pgfplots/colormap/tolYlOrBr/.style={
		colormap={tolYlOrBr}{
			color=(tolYlOrBr0);
			color=(tolYlOrBr1);
			color=(tolYlOrBr2);
			color=(tolYlOrBr3);
			color=(tolYlOrBr4);
			color=(tolYlOrBr5);
			color=(tolYlOrBr6);
			color=(tolYlOrBr7);
			color=(tolYlOrBr8);
		},
		cycle list/.define={tolYlOrBr}{[of colormap=tolYlOrBr]},
	},
	/pgfplots/cycle list/tolYlOrBr/.style={/pgfplots/@@brewer set cycle list={tolYlOrBr}},
	/pgfplots/colormap/tolIridescent/.style={
		colormap={tolIridescent}{
			color=(tolIridescent0);
			color=(tolIridescent1);
			color=(tolIridescent2);
			color=(tolIridescent3);
			color=(tolIridescent4);
			color=(tolIridescent5);
			color=(tolIridescent6);
			color=(tolIridescent7);
			color=(tolIridescent8);
			color=(tolIridescent9);
			color=(tolIridescent10);
			color=(tolIridescent11);
			color=(tolIridescent12);
			color=(tolIridescent13);
			color=(tolIridescent14);
			color=(tolIridescent15);
			color=(tolIridescent16);
			color=(tolIridescent17);
			color=(tolIridescent18);
			color=(tolIridescent19);
			color=(tolIridescent20);
			color=(tolIridescent21);
			color=(tolIridescent22);
		},
		cycle list/.define={tolIridescent}{[of colormap=tolIridescent]},
	},
	/pgfplots/cycle list/tolIridescent/.style={/pgfplots/@@brewer set cycle list={tolIridescent}},
	/pgfplots/colormap/tolRainbowWhBr/.style={
		colormap={tolRainbowWhBr}{
			color=(tolRainbowWhBr0);
			color=(tolRainbowWhBr1);
			color=(tolRainbowWhBr2);
			color=(tolRainbowWhBr3);
			color=(tolRainbowWhBr4);
			color=(tolRainbowWhBr5);
			color=(tolRainbowWhBr6);
			color=(tolRainbowWhBr7);
			color=(tolRainbowWhBr8);
			color=(tolRainbowWhBr9);
			color=(tolRainbowWhBr10);
			color=(tolRainbowWhBr11);
			color=(tolRainbowWhBr12);
			color=(tolRainbowWhBr13);
			color=(tolRainbowWhBr14);
			color=(tolRainbowWhBr15);
			color=(tolRainbowWhBr16);
			color=(tolRainbowWhBr17);
			color=(tolRainbowWhBr18);
			color=(tolRainbowWhBr19);
			color=(tolRainbowWhBr20);
			color=(tolRainbowWhBr21);
			color=(tolRainbowWhBr22);
			color=(tolRainbowWhBr23);
			color=(tolRainbowWhBr24);
			color=(tolRainbowWhBr25);
			color=(tolRainbowWhBr26);
			color=(tolRainbowWhBr27);
			color=(tolRainbowWhBr28);
			color=(tolRainbowWhBr29);
			color=(tolRainbowWhBr30);
			color=(tolRainbowWhBr31);
			color=(tolRainbowWhBr32);
			color=(tolRainbowWhBr33);
		},
		cycle list/.define={tolRainbowWhBr}{[of colormap=tolRainbowWhBr]},
	},
	/pgfplots/cycle list/tolRainbowWhBr/.style={/pgfplots/@@brewer set cycle list={tolRainbowWhBr}},
	/pgfplots/colormap/tolRainbowDiscrete/.style={
		colormap={tolRainbowDiscrete}{
			color=(tolRainbowDiscrete0);
			color=(tolRainbowDiscrete1);
			color=(tolRainbowDiscrete2);
			color=(tolRainbowDiscrete3);
			color=(tolRainbowDiscrete4);
			color=(tolRainbowDiscrete5);
			color=(tolRainbowDiscrete6);
			color=(tolRainbowDiscrete7);
			color=(tolRainbowDiscrete8);
			color=(tolRainbowDiscrete9);
			color=(tolRainbowDiscrete10);
			color=(tolRainbowDiscrete11);
			color=(tolRainbowDiscrete12);
			color=(tolRainbowDiscrete13);
			color=(tolRainbowDiscrete14);
			color=(tolRainbowDiscrete15);
			color=(tolRainbowDiscrete16);
			color=(tolRainbowDiscrete17);
			color=(tolRainbowDiscrete18);
			color=(tolRainbowDiscrete19);
			color=(tolRainbowDiscrete20);
			color=(tolRainbowDiscrete21);
			color=(tolRainbowDiscrete22);
			color=(tolRainbowDiscrete23);
			color=(tolRainbowDiscrete24);
			color=(tolRainbowDiscrete25);
			color=(tolRainbowDiscrete26);
			color=(tolRainbowDiscrete27);
			color=(tolRainbowDiscrete28);
		},
		cycle list/.define={tolRainbowDiscrete}{[of colormap=tolRainbowDiscrete]},
	},
	/pgfplots/cycle list/tolRainbowDiscrete/.style={/pgfplots/@@brewer set cycle list={tolRainbowDiscrete}},
}%

\usepackage{listings}

\lstdefinestyle{C_style}{
	basicstyle=\fontsize{9pt}{0.8em}\ttfamily\color{tolBlack}, 
	breakatwhitespace=false, 
	breaklines=true, 
	captionpos=b, 
	commentstyle=\usefont{T1}{pcr}{bm}{sl}\color{tolDarkGrey}, 
	deletekeywords={}, 
	escapeinside={*@}{@*}, 
	firstnumber=1, 
	frame=single, 
	morekeywords=[1]{
	uint8_t,uint16_t,uint32_t,uint64_t,
	int8_t,int16_t,int32_t,int64_t
	},
	keywordstyle=[1]\bfseries\color{tolDarkBlue}, 
	morecomment=[l][\bfseries\color{tolDarkRed}]{\#},
	numbers=left, 
	numbersep=10pt, 
	numberstyle=\tiny\color{tolDarkGrey}, 
	rulecolor=\color{tolBlack}, 
	showstringspaces=false, 
	showtabs=false, 
	stepnumber=5, 
	stringstyle=\color{tolDarkYellow}, 
	tabsize=3, 
}

\newcommand{\textcmd}[1]{ \colorbox{tolGrey!13!white}{\texttt{#1}}}

\begin{document}

\title{Quality tetrahedral mesh generation with HXT}

\author[1]{Célestin Marot*}
\author[1]{Jean-François Remacle}

\address[1]{
\orgdiv{Institute of Mechanics, Materials and Civil Engineering},
\orgname{Université catholique de Louvain},
\orgaddress{\state{Louvain-la-Neuve}, \country{Belgium}}
}

\abstract[Summary]{
We proposed, in a recent paper \cite{marot_one_nodate},
a fast 3D parallel Delaunay kernel
for tetrahedral mesh generation. 
This kernel was however incomplete in the sense that 
it lacked the necessary mesh improvement tools. 
The present paper builds on that previous work
and proposes a fast parallel \emph{mesh improvement} stage
that delivers high-quality tetrahedral meshes
compared to alternative open-source mesh generators.
Our mesh improvement toolkit includes
edge removal and improved Laplacian smoothing as well as a brand new operator
called the Growing SPR Cavity,
which can be regarded as the \emph{mother of all flips}. 
The paper describes the workflow of the new mesh improvement schedule, 
as well as the details of the implementation. 
The result of this research is a series of open-source scalable software components, 
called HXT,
whose overall efficiency is demonstrated on practical examples
by means of a detailed comparative benchmark
with two open-source mesh generators: Gmsh and TetGen.}

\keywords{HXT, Growing SPR Cavity, quality, tetrahedral, mesh, improvement}

\corres{
Célestin Marot,
Institute of Mechanics, Materials and Civil Engineering,\\
Université catholique de Louvain,\\
Avenue Georges Lemaitre 4, bte L4.05.02,
1348 Louvain-la-Neuve,
Belgium.\\
\email{celestin.marot@uclouvain.be}
}

\fundingInfo{
\fundingAgency{European Union's Horizon 2020 research and innovation programme}
\fundingNumber{ERC-2015AdG-694020}
}

\maketitle

\section{Introduction}

Tetrahedral meshes are the geometrical support
for most finite element discretizations. 
The size and the shape of the generated tetrahedral elements
must however be controlled cautiously 
to ensure reliable numerical simulations in industrial applications. 
The majority of popular
tetrahedral mesh generators are based on a Delaunay kernel,
because Delaunay-based algorithms are fast, especially in 3D.
They are also robust and consume relatively little memory. 
Yet, \emph{pure} Delaunay meshes
are known to contain near-zero volume elements, called \emph{slivers}, 
and a mesh improvement stage is mandatory
if one wishes to end up with a high-quality computational mesh.

We proposed, in a recent paper \cite{marot_one_nodate}, 
techniques to compute a Delaunay triangulation of three
billion tetrahedra in less than one minute using multiple threads. 
We also explained how to extend that algorithm 
to obtain an efficient parallel tetrahedral mesh generator. 
This 
mesh generator was however incomplete in the sense that 
it did not provide any mesh improvement process. 
The present paper builds on the 2019 paper
and proposes a \emph{mesh improvement} strategy 
that is both highly parallel and efficient, 
and generates high quality meshes.

Starting with an existing surface mesh, 
tetrahedral mesh generation 
is usually broken down into four steps. 
\begin{enumerate}
	\item \textbf{Empty Mesh}: A triangulation (tetrahedralization) 
      of the volume, based on all \emph{points} of the surface mesh 
      and with no additional interior nodes (hence the name empty mesh), 
      except for some isolated user-specified points in the interior of the volume,
      is first generated. 
	\item \textbf{Boundary Recovery}: 
      The recovery step locally
      modifies the empty mesh to match the triangular
      facets of the surface mesh as well as some possible embedded
      user-specified constrained triangles and lines. 
	\item \textbf{Refinement}: The empty
      mesh is iteratively refined by adding points in the interior of the
	  volumes while always preserving a valid mesh at each
      iteration. The refinement step ends up when all tetrahedra are
      smaller in size than a user-prescribed sizemap.
	\item \textbf{Improvement}: Whereas the refinement step controls the
      size of the generated tetrahedra, the improvement step finalizes the mesh by
      locally eliminating badly shaped tetrahedra, and optimizes
      the quality of the mesh by means of specific topological operations and vertex
      relocations.
\end{enumerate}

Steps 1 and 3 were described in detail in our 2019 paper\cite{marot_one_nodate}.  
Step 2 is currently done using TetGen's
implementation\cite{DBLP:journals/toms/Si15}. 
Finally, Step 4, which is the central topic in this paper,
relies on a novel mesh improvement strategy
based on an innovative local mesh modification operator 
called \emph{Growing SPR Cavity}.  
The overall efficiency of the resulting 
full-fledged open-source tetrahedral mesh generation library 
is then discussed in detail. 
Some implementation details are included in this paper, 
be the authors are  aware that 
the devil is very much into the details. 
Being commited to provide \emph{reproducible research},
the code and meshes used in this paper's test cases are thus all made available at
\url{https://git.immc.ucl.ac.be/marotc/tetmesher_benchmark}.

The computer code supporting this paper 
is available in 
Gmsh 4.6.0 \footnote{\url{https://gitlab.onelab.info/gmsh/gmsh/-/tree/master/contrib/hxt}} 
and above versions through the \textcmd{-algo hxt} parameter or the \textcmd{Mesh.Algorithm3D=10}
option. 
HXT successfully passes all Gmsh 3D benchmarks and, 
as discussed below, is faster and generates meshes of higher quality than
other open-source implementations, including Gmsh's native 3D mesher.  

The paper is structured as follows. 
Section~\ref{sec:1} contains a concise review of existing mesh improvement
operations. Whenever our implementation differs from others, the differences are
briefly explained. Section~\ref{sec:2} is devoted to the presentation 
of our new \emph{Growing SPR Cavity} operation, 
which is the main contribution of this paper.
In section~\ref{sec:3}, we detail our mesh improvement schedule, its parallelization and
the quality improvement that can be obtained. Finally, in Section~\ref{sec:4}, we present
our complete tetrahedral mesh generator algorithm (HXT) 
and compare it against Gmsh and TetGen,
two other reference
open-source tetrahedral mesh generation softwares. 

\section{Mesh improvement operations}\label{sec:1}

\emph{Mesh improvement operations} are  modifications of the mesh aiming at increasing
its overall quality, the latter being essentially determined by the quality of its
worst quality element\cite{shewchuk_what_2002}. Therefore, when we shall speak
of the quality of a mesh or of a cavity, we shall in practice refer to
the quality of its worst tetrahedron. 
The value of a quality measure is expected to be
inversely proportional to the interpolation error 
associated with the tetrahedron in some discretization scheme. 
If the tetrahedron is not valid, i.e., if it is inverted or flat,
the quality measure should then be non-positive. 
Two examples of quality measure, the Gamma and SICN measure, are
described in Appendix \ref{bench:gamma} and \ref{bench:sicn}.

\begin{figure}[htb]
\centering
\begin{subfigure}[b]{0.48\textwidth}
	\centering
	\includegraphics[width=0.95\textwidth]{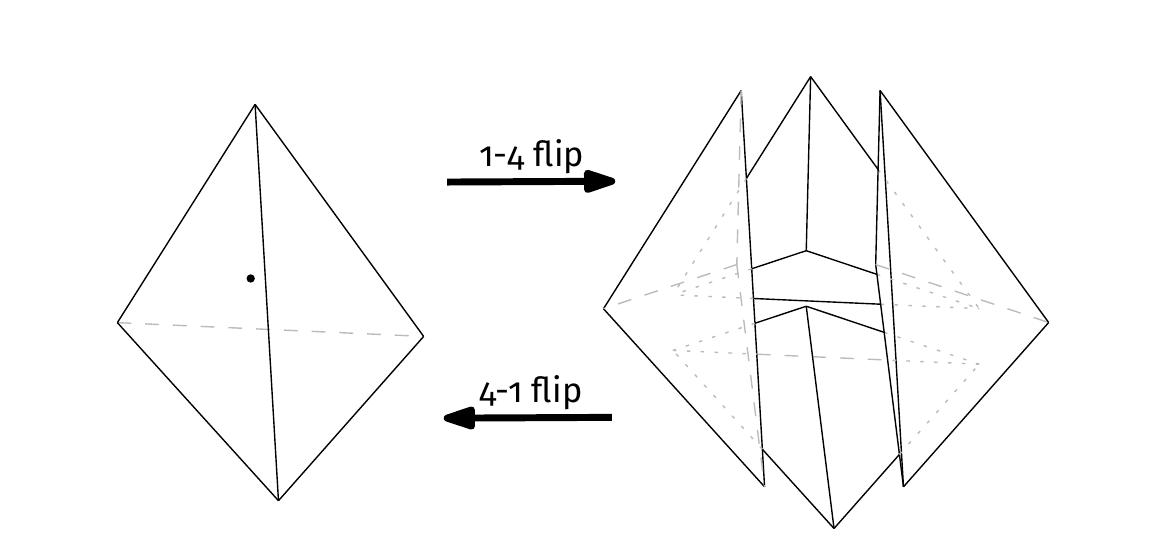}
	\caption{The 1-4 move (point insertion) and 4-1 move (point removal)}
	\label{fig:flip14}
\end{subfigure}
	~
\begin{subfigure}[b]{0.48\textwidth}
\centering
\includegraphics[width=0.95\textwidth]{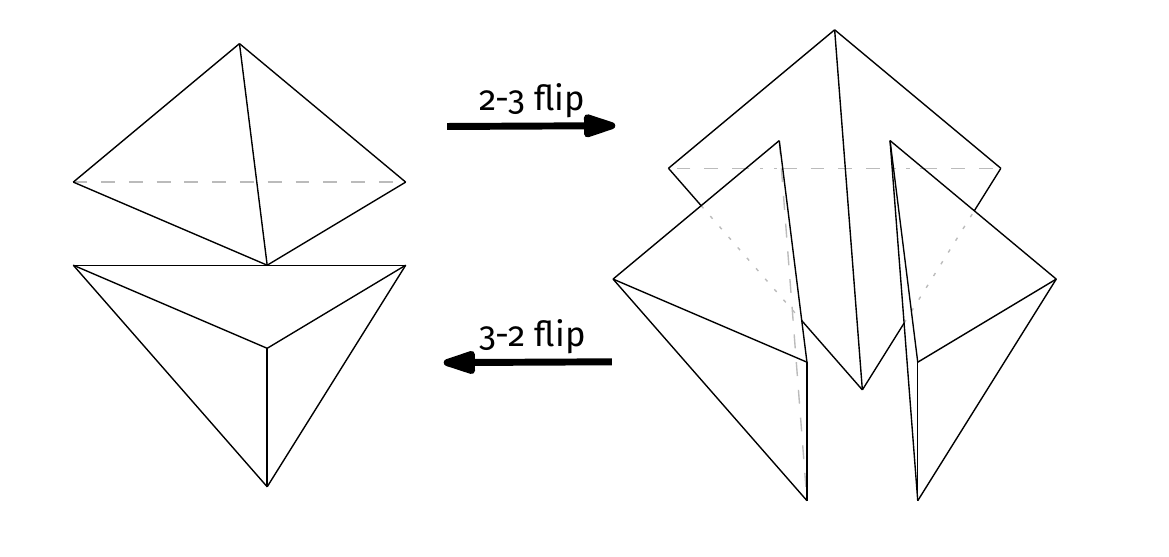}
\caption{The 2-3 and 3-2 flips}
\label{fig:flip23}
\end{subfigure}
\caption{Pachner moves}
\end{figure}

A cavity is a volume corresponding to a set of face-connected tetrahedra. 
A mesh modification operates on a cavity, 
which is by definition the part of the mesh that is modified. 
Among all possible mesh modifications, 
mesh improvement operations are those 
that strictly increase the quality of a cavity, 
from an original tetrahedralization of quality $q_a$ to an  improved
tetrahedralization of quality $q_b > q_a$. 
Any mesh modification can be decomposed into a succession of elementary moves, 
called bistellar flips or Pachner moves\cite{pachner_pl_1991}. 
The 1-4 move adds a point inside a cavity formed by
the removal of a single tetrahedron, and fills it with 4 new tetrahedra. 
The 4-1 move is the opposite, it removes a point. 
Both operations are illustrated on Figure \ref{fig:flip14}. 
The two remaining Pachner moves are the 2-3 and 3-2 flips,
which do neither add nor remove any points in the
tetrahedralization (see Figure \ref{fig:flip23}).

\subsection{Flipping}

\begin{figure}[htb]
\centering
\begin{subfigure}[b]{0.48\textwidth}
	\centering
	\includegraphics[width=0.95\textwidth]{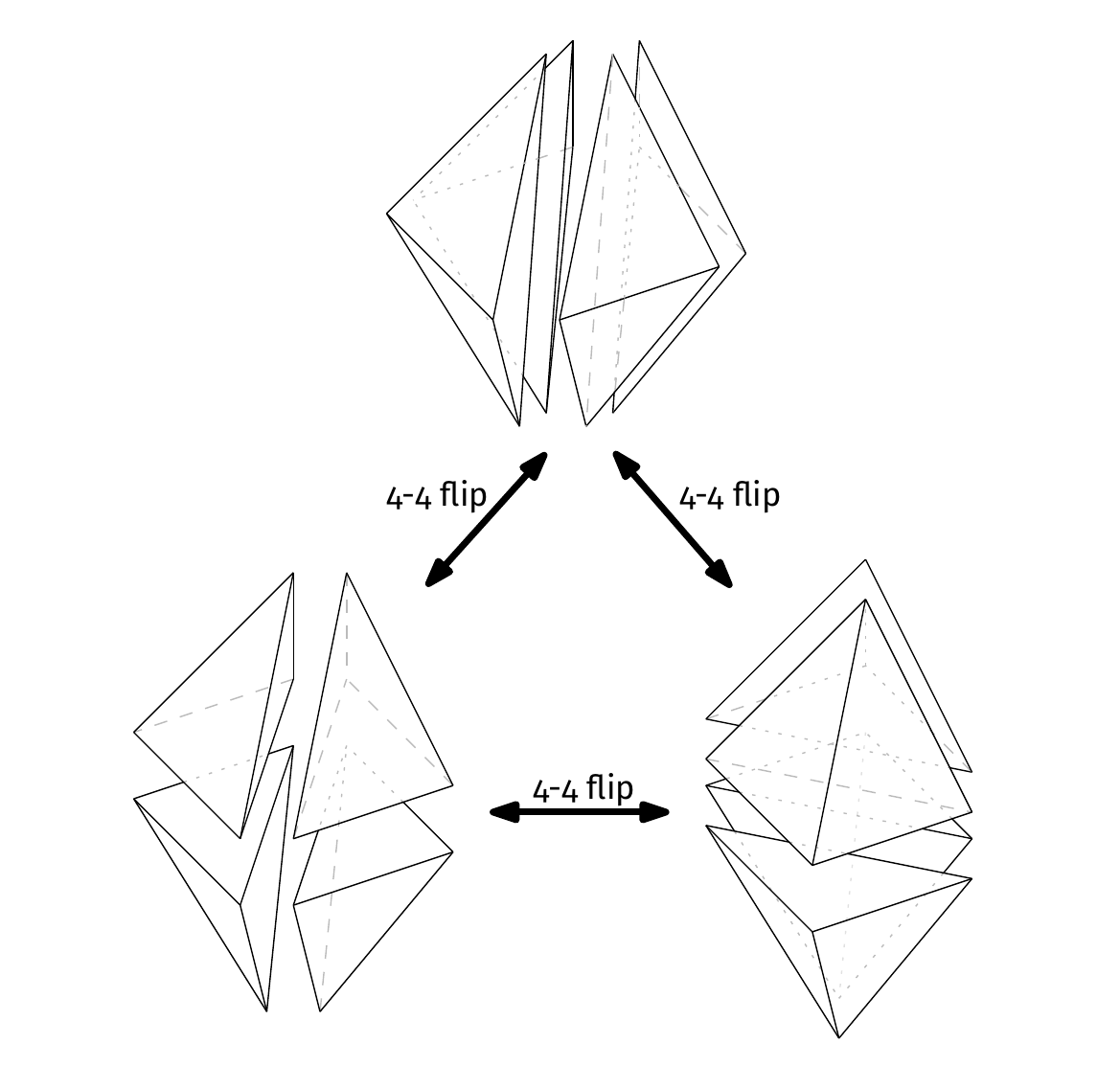}
	~\\[1cm]
	\caption{The 4-4 flip}
	\label{fig:flip44}
\end{subfigure}
~
\begin{subfigure}[b]{0.48\textwidth}
	\centering
	\includegraphics[width=0.95\textwidth]{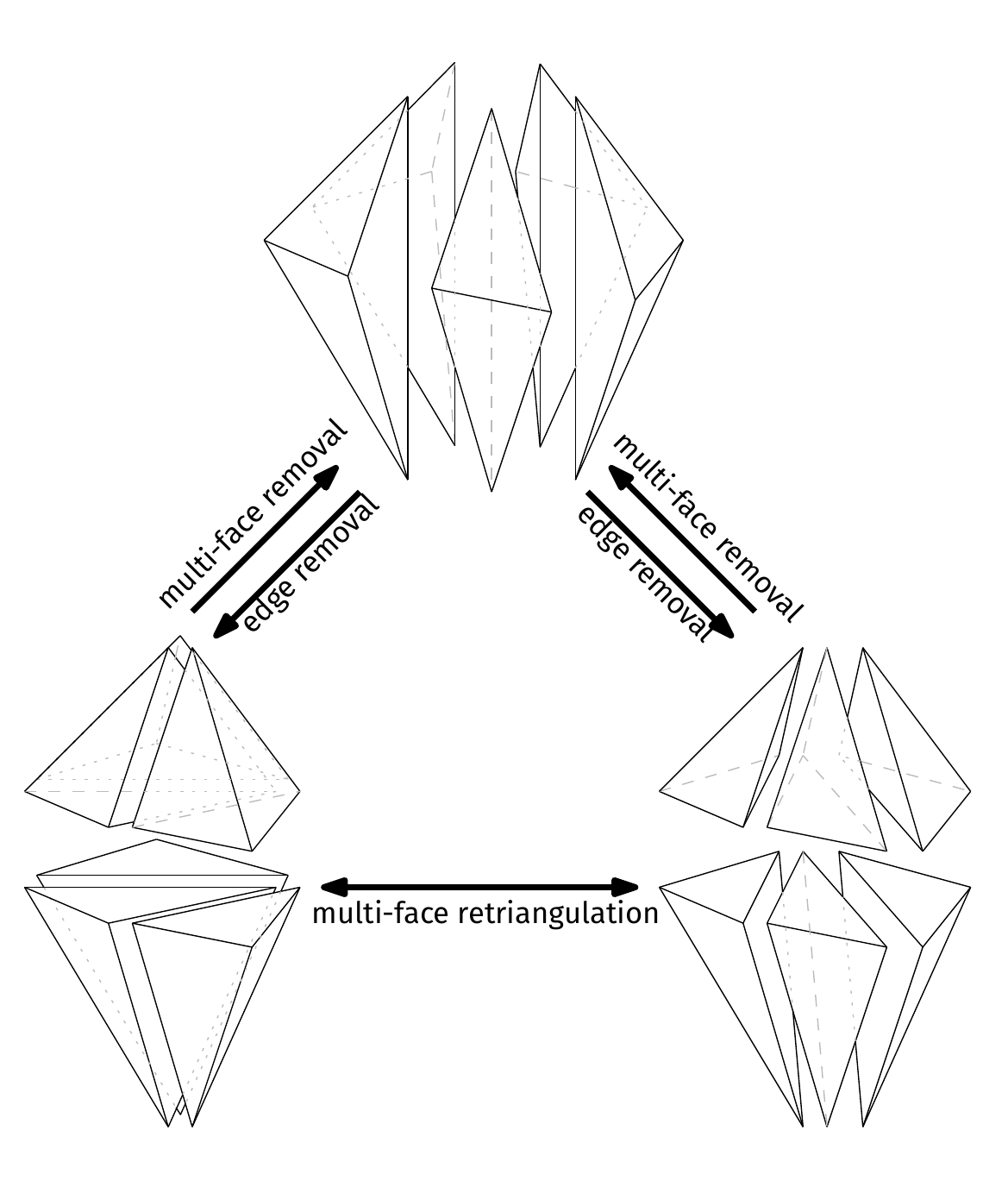}
	\caption{Edge removal, multi-face removal and multi-face retriangulation}
	\label{fig:multiplus}
\end{subfigure}
\caption{Examples of composite topological operations}
\end{figure}

The purpose of the mesh refinement step 
is to obtain a distribution of points all over the volume 
whose distance to the closest neighbour 
is prescribed by a given mesh size field.
Adding or removing vertices during the subsequent
mesh improvement phase 
is therefore likely to disrupt the refined point distribution
and see it deviate from the prescribed size map. 
The simplest mesh improvement schedule 
one may think of 
is thus composed of 2-3 and 3-2 flips only,
the simplest operations that do not add or remove points,
and are executed whenever they improve the quality 
of the tetrahedralization in their respective cavity. 
This strategy is able to eliminate efficiently most ill-shaped tetrahedra.
However, this hill-climbing method often reaches local maxima
where 2-3 or 3-2 flip no longer improve the mesh quality
although the overall quality is not yet optimal.
To overcome this limitation, combinations of multiple flips can be applied at
once whenever one can check they result in a tetrahedralization of better
quality.
This is equivalent with creating more complex topological operations. 
For example, the 4-4 flip, which is an operation on a cavity of 6 points and 4 tetrahedra
with only one interior edge (see Figure \ref{fig:flip44}),
can be obtained by doing a 2-3 flip that creates a new interior
edge, followed by a 3-2 flip that removes the initial edge.

\subsection{Edge removal}

The most useful topological operation is 
the \emph{edge removal} operation. 
It is a generalization of the 3-2 and 4-4 flips,
starting from a cavity with $N\ge3$ tetrahedra surrounding 
a unique edge $\overline{ab}$
(see Figure \ref{fig:multiplus}). 
The operation removes first all tetrahedra adjacent to that edge,
and creates instead $N-2$ \emph{upper} tetrahedra connected to $a$ 
and $N-2$ \emph{lower} tetrahedra connected to $b$. 
The facets between \emph{lower} and \emph{upper} tetrahedra 
form a 2D \emph{sandwiched} triangulation,
and Shewchuk has proposed an algorithm, 
based on dynamic programming concepts, 
that finds the 2D triangulation resulting in an optimal 
tetrahedralization\cite{shewchuk_two_nodate}. 
In contrast, Si presents a more
versatile implementation of the edge removal operation
that recursively removes other edges that prevents the
current edge removal from producing a valid
tetrahedralization\cite{DBLP:journals/toms/Si15}.
This forms a tree of edge removal operations, where the leafs are removed
with a sequence of 2-3 flips terminated by a final 3-2 flip.
Our implementation of edge removal does neither use dynamic programming
nor sequences of flips. It is rather a brute-force approach inspired by the
first description of edge removal by Freitag and
Ollivier-Gooch\cite{freitag_tetrahedral_1997}. 
In a nutshell, the principle is as follows.
Each triangle of the sandwiched 2D triangulation is assigned a quality, 
which is the minimum of the corresponding upper and lower tetrahedron quality. 
If the quality of a triangle is less than the quality of the original
tetrahedralization, the triangle is marked bad. 
Using precomputed
tables giving all possible triangulations up to $N=7$
in which this triangle is found, 
it is then possible to eliminate triangulations 
involving bad triangles.
Whenever all triangulations are eliminated, 
the edge removal is not performed. 
If, on the other hand, several tetrahedralizations are possible, 
their respective overall quality are computed, and the best candidate is selected. 
As Freitag and Ollivier-Gooch noted, 
having more than 7 tetrahedra around an edge 
is exceptional,
and edge removal is also less likely to succeed in those cases.
Our approach favors thus simplicity over asymptotic complexity,
although all approaches have of course their pros and cons in practice.
In average, our edge removal terminates within about 1 microsecond
for cavities of about 5 tetrahedra, on modern hardware.

The inverse of the edge removal was coined \emph{multi-face removal} by
Shewchuk\cite{shewchuk_two_nodate}. The operations that derive from edge
removal are shown in Figure \ref{fig:multiplus}. This group of operations
includes a third operation, named \emph{Mutli-face retriangulation} by Misztal et
al.\cite{clark_tetrahedral_2009}. Mutli-face retriangulation can be regarded
either as a combination of an edge removal and a multi-face removal, or as a
sequence of 4-4 flips modifying the 2D sandwiched triangulation. Although
multi-face removal and multi-face retriangulation have proved their
effectiveness, their implementation is more involved than edge removal, and they
are far less used. We decided not to implement them, because they are covered by
the \emph{Growing SPR Cavity} operation anyway. This \textit{mother of all
flips} is detailed in section \ref{sec:2}.

\subsection{Smoothing}

A vertex relocation, or smoothing operation, 
in the context of tetrahedral mesh optimization,
is an operation that changes the position of a point in order to
improve the quality of the adjacent tetrahedra. Smoothing methods have been
extensively studied in the past 25 years\cite{amenta_optimal_1999,
freitag_parallel_1999, freitag_tetrahedral_2002, dassi_tetrahedral_2018},
and their objective is twofold: improve the overall quality of the mesh and
space out points appropriately 
so that subsequent topological transformations can further improve the mesh. 
Smoothing and topological transformations are indeed
more effective when combined \cite{freitag_tetrahedral_1997}. The most used smoothing
technique, called \emph{Laplacian smoothing}, simply relocates a
point at the centroid of the set of points to which is it connected by a mesh edge.
As for flipping, Laplacian smoothing is applied only if it effectively improves the
quality of the mesh. In our mesh generation library \emph{HXT},
Laplacian smoothing is combined with a golden-section search
of the optimal relocation on the segment 
between the original position and the centroid.
This approach is effective in practice,
although the objective function may have local maxima over the segment, 
and the golden-section search is not guaranteed to identify the largest one. 
In their work\cite{freitag_parallel_1999}, Freitag et al. studied
this optimized Laplacian smoothing technique, among other techniques.

\subsection{Other operations}

A good review of classical mesh improvement operations is found in
Klingner's Ph.D. Dissertation\cite{Klingner_improving_2008}.
Klingner reports therein
that large quality improvements are obtained by using point insertion and edge
contraction. We have not implement those operations so far, 
but we will consider them for later updates. 
They are indeed more complex than one might think at first sight. 
It is not enough to insert or remove points and check
on whether the quality is improved. 
The position of adjacent points must also often be modified 
to obtain tetrahedra of better quality and,
more importantly, to respect the meshsize map.

\begin{figure}[htb]
\centering
\includegraphics[width=0.5\textwidth]{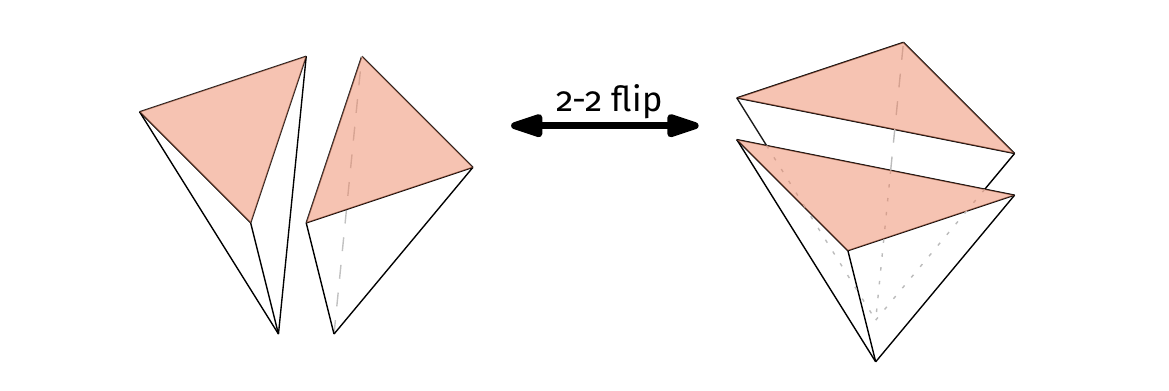}
\caption{The 2-2 flip. Triangles of the surface mesh are shaded in orange.}
\label{fig:flip22}
\end{figure}

Valid mesh of higher quality may be obtained in some situations 
if a slight modification of the surface mesh is allowed.
This however implies relying on a CAD software during the meshing process
to evaluate the distance to the parametric definition of the initial surface.
A more lightweight approach considering a modification of the surface
mesh itself, up to an approximated Hausdorff distance threshold,
would still mean bookkeeping in memory 
an unmodified version of the surface mesh. 
As we are aiming at a simple and parallelizable mesh generator,  
we chose to not consider at all operations that modifies the surface mesh. 
In some cases, this is even an asset,
as there are situations where one wants the surface mesh to remain
strictly unchanged, for instance, if the surface is an interface 
between independent parts of an assembly.
Boundary modification is however an option worth being considered
for future implementations, as it has shown its
effectiveness\cite{brewer_aggressive_2008}. 
The most elementary surface mesh modification operation
simply flips a pair of adjacent tetrahedra, which are themselves
adjacent to two boundary triangles on the surface mesh. This operation is called
a \emph{2-2 flip} and is illustrated in Figure \ref{fig:flip22}.

\section{Growing SPR cavity}\label{sec:2}

\begin{figure}[htb]
	\centering
	\includegraphics[scale=0.5]{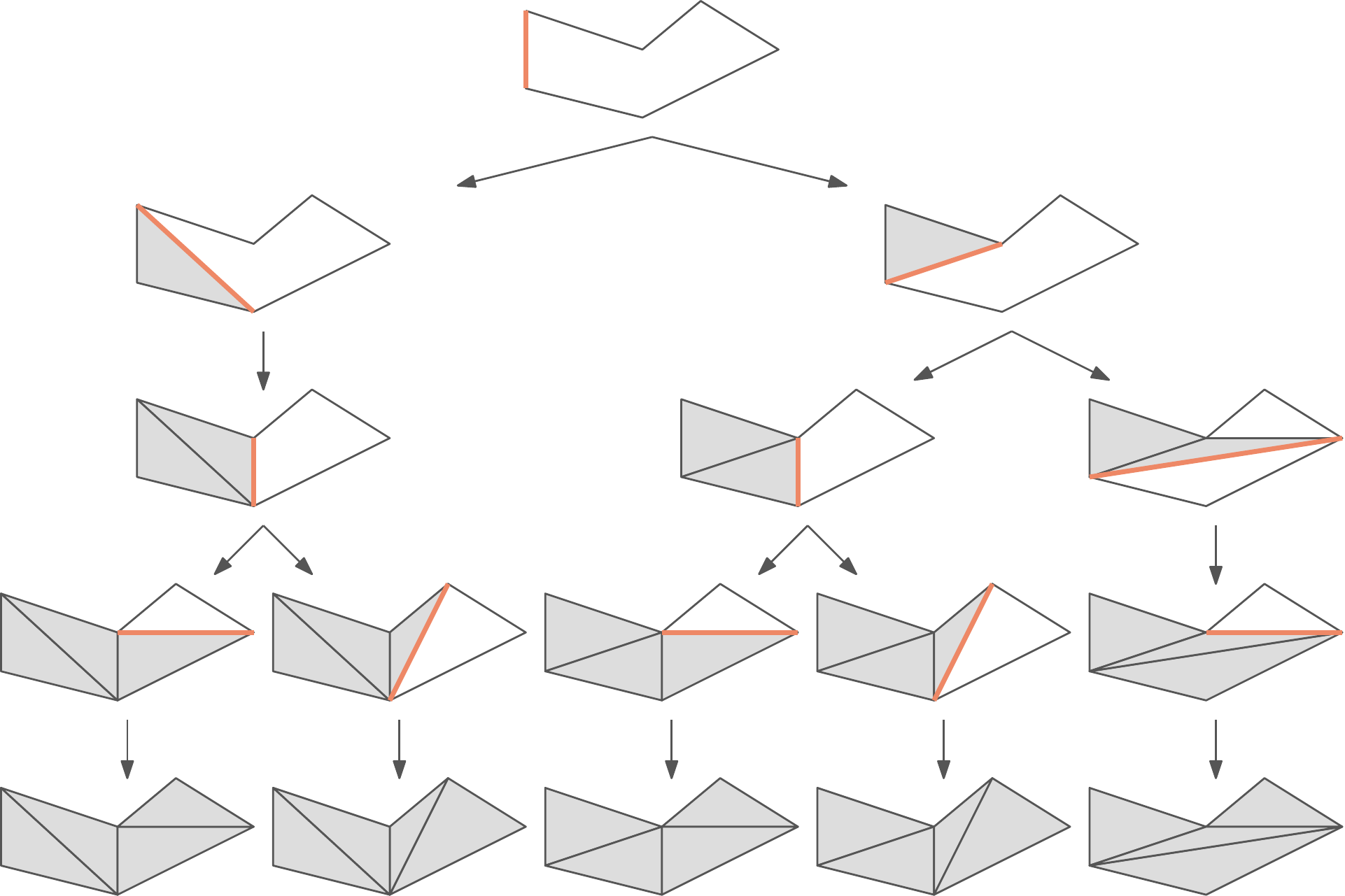}
	\caption{2D illustration of the SPR search tree}
	\label{fig:SPR}
\end{figure}

\emph{Small Polyhedron Reconnection} (SPR) is a branch and bound type algorithm
for finding the best of all possible tetrahedralization of a cavity. 
It was named and first implemented by Liu et al. in 2006\cite{liu_small_2009}. 
Finding whether a polyhedron can be triangulated or
not is already a NP-complete problem\cite{ruppert_difficulty_1992}.
Finding the best triangulation is an even more difficult NP-hard problem. The SPR
algorithm has indeed factorial complexity with regard to the number of points
$n$ in the cavity. It starts from a selected face, and recursively fills up the
cavity with well-shaped tetrahedra, 
trying out each of the $n-3$ remaining points, 
until the best
triangulation is found. The basic algorithm thus covers the whole tree
of possible triangulations, but simple optimizations allow pruning most
branches\cite{liu_small_2009, marot_reviving_2020}. Figure \ref{fig:SPR} shows
the search tree for a simple 2D cavity.

The choice of the cavity on which to apply the SPR operation 
is a matter that received very little attention, 
although the performances strongly depend 
on the geometry and topology of the cavity. In this paper, we
propose a new algorithm called \emph{Growing SPR Cavity} (GSC),
whose basic idea is
to increment the number of points gradually, and to apply the SPR at each step
until a better triangulation is found. As the SPR algorithm has factorial
complexity, the cost of repeating SPR operations from 4 to $n$ points is
not much higher than computing directly the best triangulation with $n$ points. In addition, most of the SPR structure can be reused from one iteration
to the next, and a better triangulation of the cavity is usually found within few
iterations. The limit to the maximum number of points has been set to $n=32$,
and GSC abandons if no better triangulation has been found 
when the cavity has reached 32 points. 

We explained in \cite{marot_reviving_2020} how the SPR algorithm can be
optimized by storing, in a $n^4$ table, the quality (and thus also the validity)
of all tetrahedra the algorithm already encountered. By limiting the number of
points to 32, we can allocate an acceptably large $32^4$ table at the beginning
of the GSC algorithm and keep it from one iteration of the GSC algorithm to
another.


\begin{figure}[htb]
\centering
\begin{subfigure}[b]{0.2\textwidth}
\centering
\includegraphics[page=1, width=\textwidth]{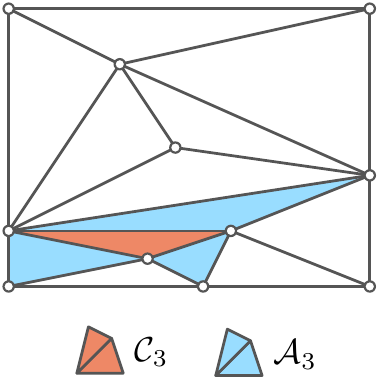}
\caption{}
\end{subfigure}
\hspace{0.04\textwidth}
\begin{subfigure}[b]{0.2\textwidth}
\centering
\includegraphics[page=2, width=\textwidth]{images/GSC}
\caption{}
\end{subfigure}
\hspace{0.04\textwidth}
\begin{subfigure}[b]{0.2\textwidth}
\centering
\includegraphics[page=3, width=\textwidth]{images/GSC}
\caption{}
\end{subfigure}
\hspace{0.04\textwidth}
\begin{subfigure}[b]{0.2\textwidth}
\centering
\includegraphics[page=4, width=\textwidth]{images/GSC}
\caption{}
\end{subfigure}
\caption{Growing SPR Cavity in 2D. Triangles in the cavity ${\mathcal
C}$ are colored in orange, and triangles 
adjacent to the cavity
${\mathcal A}$ are colored in cyan. 
\textbf{(a)}: at first, the cavity only contains a bad triangle. 
\textbf{(b)} and \textbf{(c)}: every point outside the cavity is
a vertex of at most one triangle in ${\mathcal A}$, 
so we add the triangle in
${\mathcal A}$ with the worst quality to ${\mathcal C}$. 
\textbf{(d)}: the SPR
algorithm found a better triangulation for ${\mathcal C}_5$, the triangulation
of the cavity is replaced and the GSC algorithm ends.}
\label{fig:gsc2d}
\end{figure}

\begin{algorithm}[htb]
\caption{The \emph{Growing SPR Cavity} (GSC) operation, applied on a bad
tetrahedron $t$ of a mesh ${\mathcal  M}$}
\label{algo:GSC}

\begin{algorithmic}[1]
\algnotext{EndIf}%
    \Function{GSC}{$t$, ${\mathcal  M}$}
    	\State ${\mathcal  C} \gets t$  \Comment the cavity is the tetrahedron at first
    	\State $n \gets 4$              \Comment the number of points in ${\mathcal  C}$
    	\While{$n<32$}
    	\State ${\mathcal  C} \gets$ \Call{ExtendCavity}{${\mathcal  M}$, $C$, $n$} \Comment add $p_{k+1}$ and all tetrahedra with 4 vertices in ${\mathcal P}_{k+1}$
    	\State $n \gets n+1$\\
    	\State ${\mathcal  C_b} \gets$ \Call{SPR}{${\mathcal  C}$} \Comment find the optimal triangulation of ${\mathcal  C}$
    	\If{${\mathcal  C_b} \ne {\mathcal  C}$}

    		\State \Return $({\mathcal  M} \setminus {\mathcal  C}) \cup {\mathcal  C_b}$ \Comment Mesh is improved
    	\EndIf
        \EndWhile\\
        \State \Return ${\mathcal  M}$ \Comment Failed to improve the mesh
    \EndFunction
\end{algorithmic}
\end{algorithm}

The GSC pseudocode given in algorithm \ref{algo:GSC} can be 
explained as follows. 
As its initial cavity, GSC starts with a tetrahedron of bad quality which needs to be optimized.
Let the four point of this initial tetrahedron
be denoted as $\{p_1,p_2, p_3, p_4\}$. 
Now, let ${\mathcal C}_k$ be  a cavity containing a set of $k$ 
points ${\mathcal P}_k =\{p_1, p_2,\ldots, p_k\}$. 
To iteratively obtain ${\mathcal C}_{k+1}$, 
a point $p_{k+1}$ is added and every tetrahedra whose four
points are in ${\mathcal P}_{k+1} = {\mathcal P}_k \cup p_{k+1}$ are also added.
The selection of the next point of $p_{k+1}$ 
is a heuristic based on a simple intuition comforted
by experience in testing the SPR algorithm. We observed that it is
in general easier to find a better tetrahedralization for a cavity with few
points and many tetrahedra than for a cavity with few tetrahedra per point.
Therefore, the point $p_{k+1}$ is chosen so as to add as many tetrahedra as possible. 
Let ${\mathcal A}_k$ denote the set
of tetrahedra
sharing at least one facet with a tetrahedron in ${\mathcal C}_k$. 
In practice, GSC adds the \emph{most connected point}, 
i.e., the point adjacent to the largest number of tetrahedra in ${\mathcal A}_k$. 
If several points are 
adjacent to $m$ tetrahedra in ${\mathcal A}_k$, 
the sum of the qualities
of those $m$ tetrahedra is evaluated for each point,
and the point with the lowest sum is selected as a tiebreaker rule.
Figure~\ref{fig:gsc2d} illustrates the GSC algorithm in 2D, 
where the tiebreaker rule has been used twice. 
This rule is however empirical and different alternative rule were also tested: 
selecting the point with the highest sum of
qualities, with the maximum or minimum quality, with the quality function
replaced by the volume or by the height associated to the boundary facet. 
The proposed tiebreaker rule consistently gave better results in our full mesh
improvement schedule, as detailed in section \ref{sec:3}.

In 2D, the triangles to be added to the cavity, i.e. with all 3 points in
${\mathcal P}_{k+1} = {\mathcal P}_k \cup p_{k+1}$, are always 
in ${\mathcal A}_k$. However, in 3D there might be tetrahedra that are
neither in ${\mathcal A}_k$ nor in ${\mathcal C}_k$, but still have all their vertices
in ${\mathcal P}_{k+1}$. Therefore, in reality, the GSC algorithm does not
always add the \emph{optimal point} that result in the addition of as many
tetrahedra as possible. Choosing the \emph{most connected point} and not the
\emph{optimal point} is however simpler and faster in practice.

\section{Mesh improvement schedule}\label{sec:3}

\begin{figure}[hbt]
	\centering
	\begin{subfigure}[b]{0.3\textwidth}
		\centering
		\includegraphics[height=5.25cm]{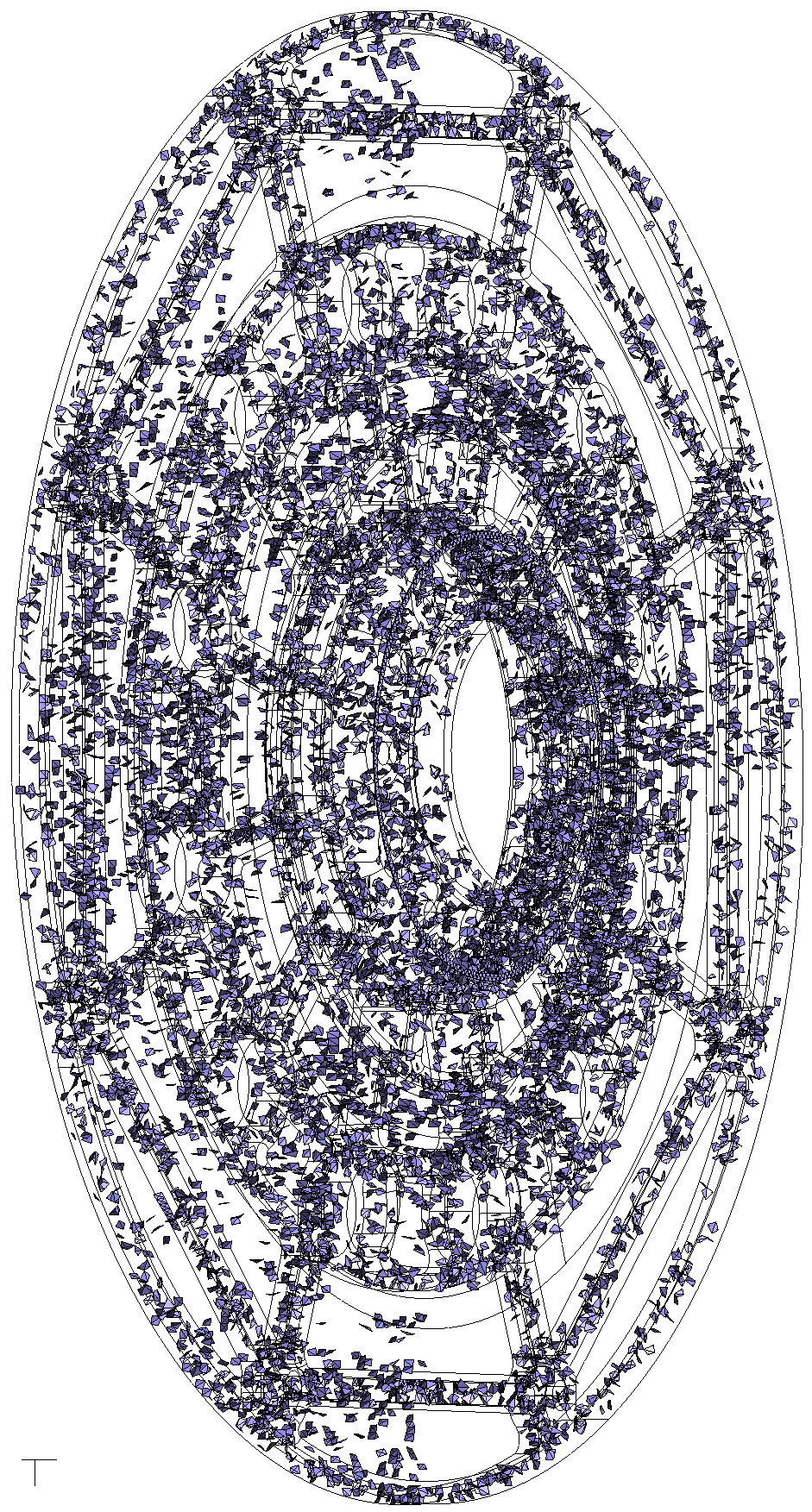}\\
		\includegraphics[height=5.25cm]{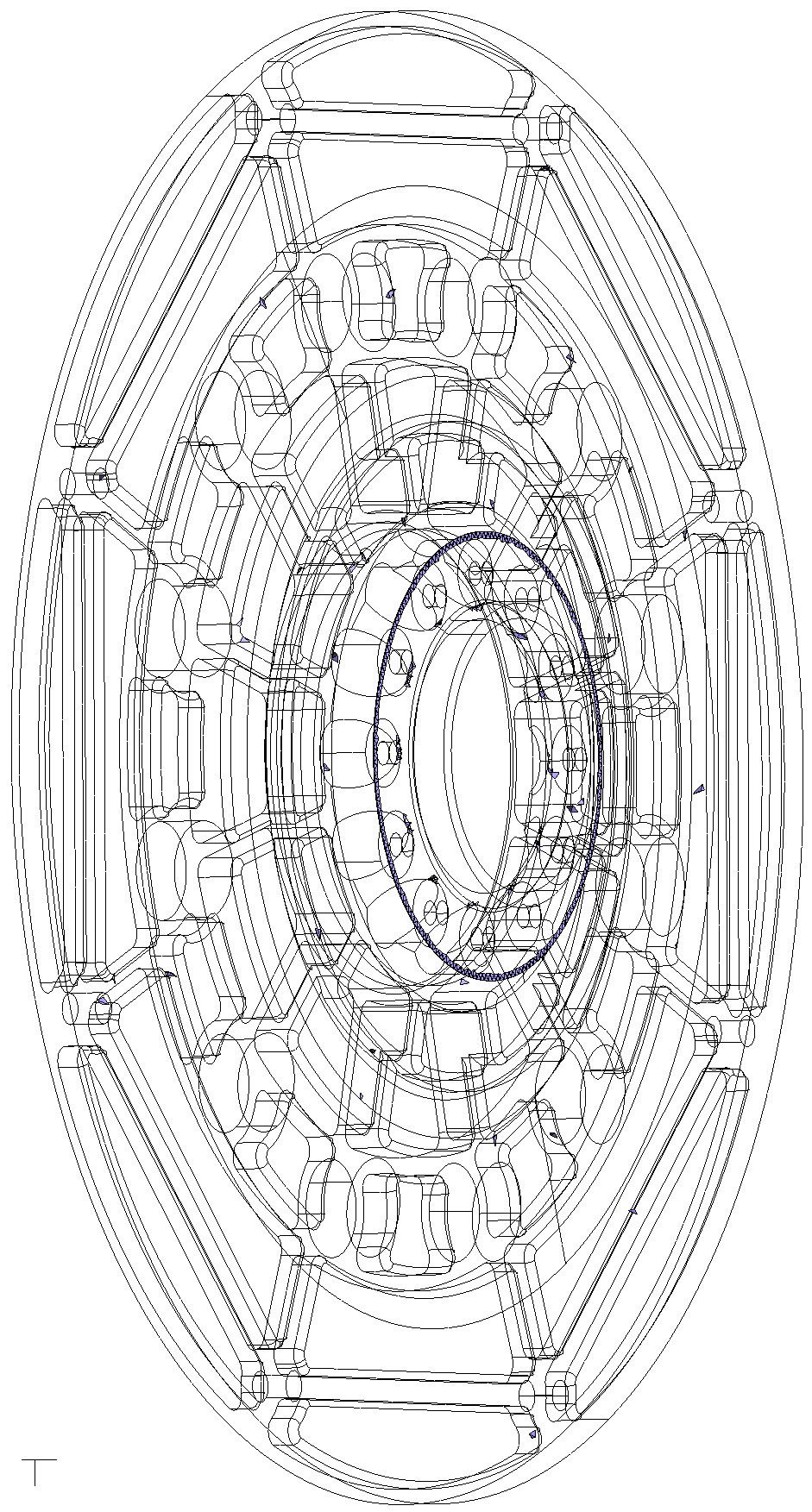}
		\caption{Tetrahedra with $\gamma < 0.35$ before (above) and after (below) the mesh improvement step on the \emph{Rotor} mesh}
	\end{subfigure}
	\hspace{0.03\textwidth}
	\begin{subfigure}[b]{0.3\textwidth}
		\centering
		\includegraphics[height=5.25cm]{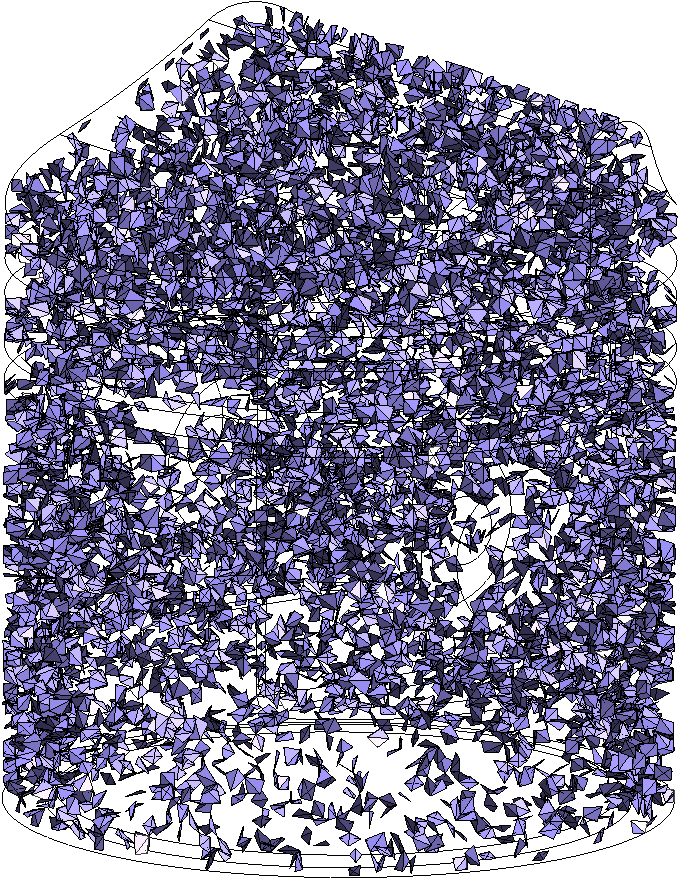}\\
		\includegraphics[height=5.25cm]{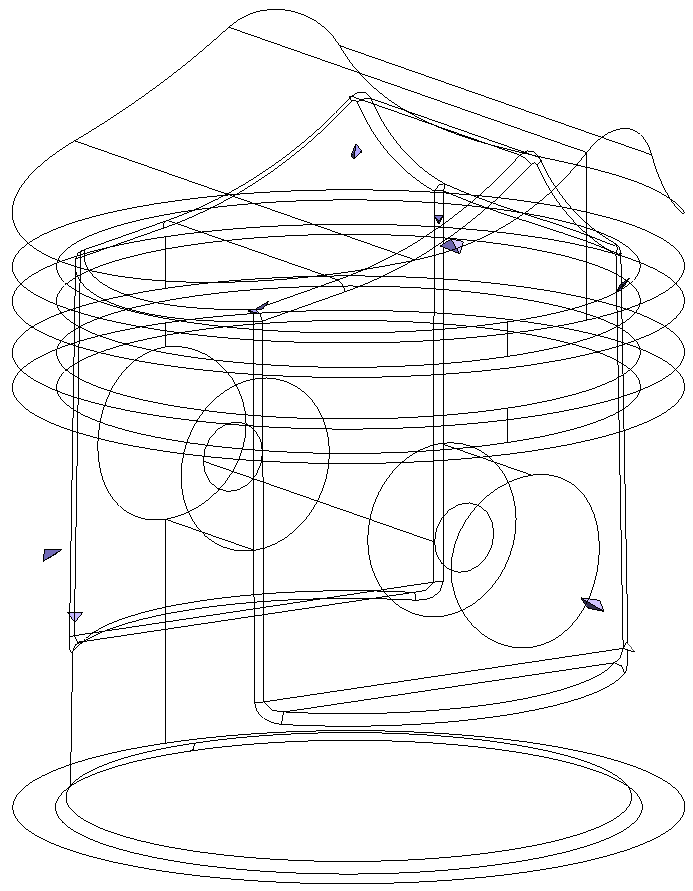}
		\caption{Tetrahedra with $\gamma < 0.35$ before (above) and after (below) the mesh improvement step on the \emph{Piston} mesh}
	\end{subfigure}
	\hspace{0.03\textwidth}
	\begin{subfigure}[b]{0.3\textwidth}
		\centering
		\includegraphics[height=5.25cm]{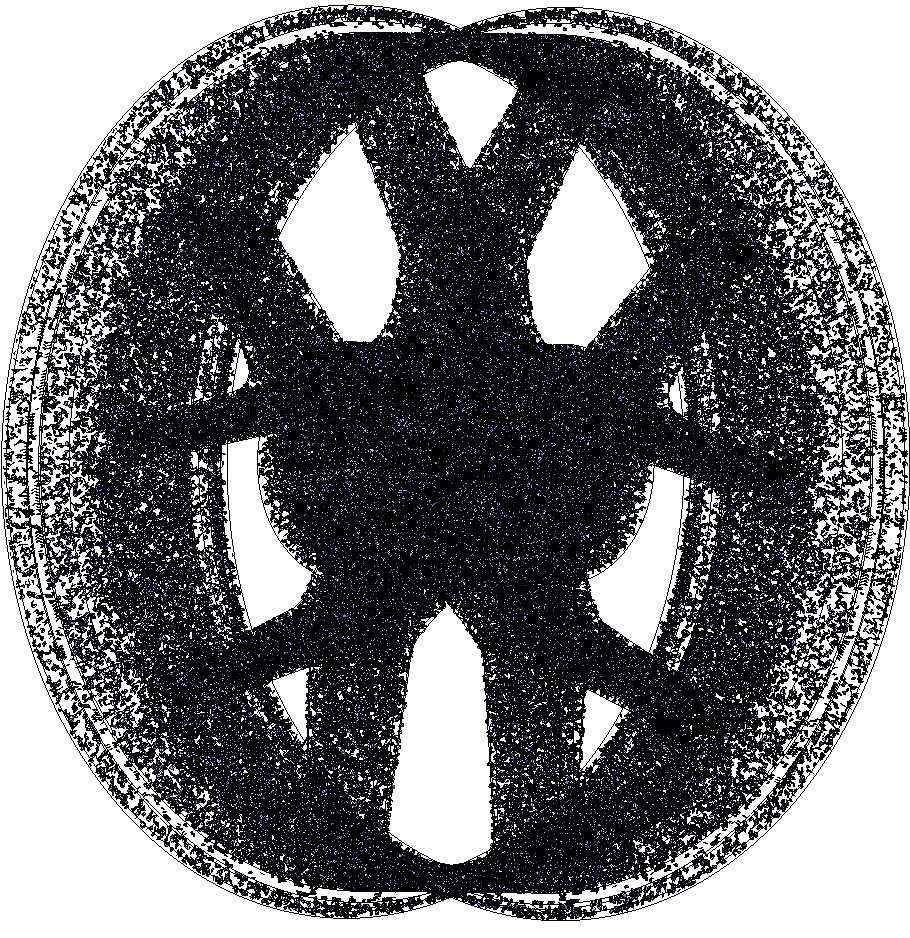}\\
		\includegraphics[height=5.25cm]{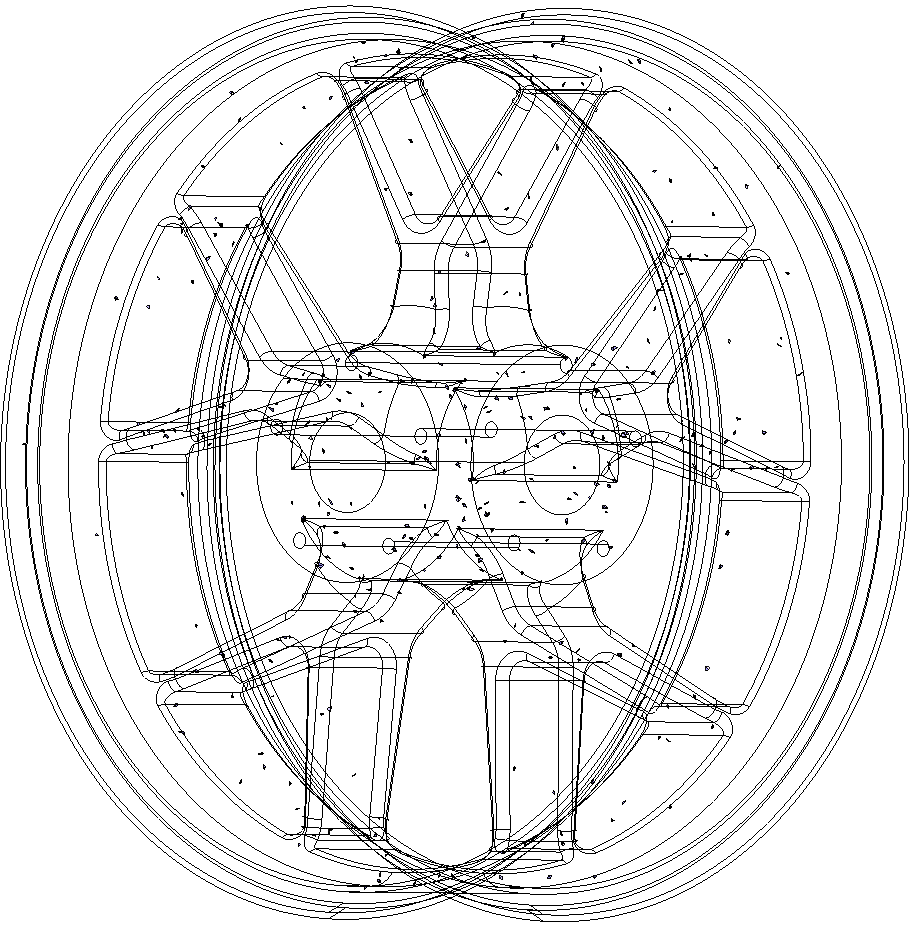}
		\caption{Tetrahedra with $\gamma < 0.35$ before (above) and after (below) the mesh improvement step on the \emph{Rim} mesh}
	\end{subfigure}
	\caption{Effect of HXT's mesh improvement step on bad tetrahedra. The
	threshold for being considered a \emph{bad} tetrahedron was set to
	$\gamma_{threshold}=0.35$. Almost all bad tetrahedra of the \emph{Rotor},
	\emph{Piston} and \emph{Rim} meshes (specified in Table \ref{tab:models})
	were improved above this threshold. }
	\label{fig:gamma035}
\end{figure}

Our mesh improvement strategy includes a Laplacian smoothing phase, an
edge removal phase and the GSC. As Laplacian smoothing and edge removal are
approximately $100\times$ faster than GSC, they are used in priority, 
whereas GSC
is used as a last resort technique to unlock processes
trapped in local maxima of the mesh quality objective function. 
The pseudocode for a simplified serial mesh improvement schedule 
is presented in Algorithm~\ref{algo:schedule}.
The \texttt{for} loop on line
\ref{algo:loopSER} will be called the \emph{SER loop} in the following
(Smoothing Edge Removal)
and the loop where the GSC operation is applied, on line \ref{algo:loopGSC}, 
will be called \emph{GSC loop}. Both loops iterate over bad tetrahedra, which are
tetrahedra with a quality under a user-defined threshold. Using the Gamma
quality function detailed in Appendix~\ref{bench:gamma}, 
our mesh improvement strategy is able to eliminate
most tetrahedra with $\gamma<0.35$, 
as shown in Figure~\ref{fig:gamma035},
where one can see that only a few bad tetrahedra subsist. 

\begin{algorithm}[!htb]
\caption{The proposed serial mesh improvement schedule 
tries to improve each tetrahedron
$\tau$ from a list of bad tetrahedra ${\mathcal T}$ in a mesh $\mathcal M$.
Bad tetrahedra are tetrahedra with a quality smaller than a user-defined
threshold $q_{min}$}
\label{algo:schedule}

\begin{algorithmic}[1]
\algdef{SE}[DOWHILE]{Do}{doWhile}{\algorithmicdo}[1]{\algorithmicwhile\ #1}%
\algnotext{EndWhile}%
\algnotext{EndIf}%
\algnotext{EndFor}%
\Function{MeshImprovement}{${\mathcal  M}$, $q_{min}$}
\Do
	\State $\text{modifGSC} \gets 0$ \Comment count modifications of the mesh by Growing SPR Cavity
	\Do
		\State $\text{modifSER} \gets 0$ \Comment count modifications of the mesh by Smoothing or Edge Removal
		\State ${\mathcal T} \gets$ \Call{GetBadTetrahedra}{${\mathcal  M}$, $q_{min}$}
	    \For{$\tau \in {\mathcal T}$} \label{algo:loopSER} \Comment the \emph{SER loop}
	        \State $\text{improved} \gets \text{\textbf{False}}$
	        \For{$\text{point} \in \tau$ \textbf{and} $\neg \text{improved}$}
	            \State $\text{improved} \gets$ \Call{LaplacianSmoothing}{${\mathcal  M}$, $\text{point}$}
	        \EndFor
	        \For{$\text{edge} \in \tau$ \textbf{and} $\neg \text{improved}$}
	            \State $\text{improved} \gets$ \Call{edgeRemoval}{${\mathcal  M}$, $\text{edge}$}
	        \EndFor
	        \If{$\text{improved}$}
	            \State $\text{modifSER} \gets \text{modifSER}+1$
	        \EndIf
	    \EndFor
	\doWhile{$\text{modifSER} > 0$}\;
	\State ${\mathcal T} \gets$ \Call{GetBadTets}{${\mathcal  M}$, $q_{min}$}
	\For{$\tau \in {\mathcal T}$} \label{algo:loopGSC} \Comment the \emph{GSC loop}
		\If{\Call{GSC}{${\mathcal  M}$, $tau$}}
			\State $\text{modifGSC} \gets \text{modifGSC}+1$
		\EndIf
	\EndFor
\doWhile{$\text{modifGSC} > 0$}\;
\EndFunction
\end{algorithmic}
\end{algorithm}

\subsection{Parallelization}

The mesh improvement strategy described above 
can be parallelized pretty much in the same way as the Delaunay refinement. 
The parallel shared-memory Delaunay kernel introduced 
in our previous article~\cite{marot_one_nodate}
partitions the domain on basis of a 3D Moore curve
defined in such a way that all partitions
contain the same amount of points to be inserted. 
A point is in a partition if its Moore index is in the range
that defines the partition. A tetrahedron, then, is considered to belong to a partition
if at least 3 of its 4 vertices lay in that partition. When the cavity created
for the insertion of a new point overlaps the boundary between different
partitions, the operation is suspended,
until all other
insertions have been tried. 
At that moment, a new Moore curve and hence new partitions are created, 
and the insertion loop resumes with the points whose insertion was suspended. 
The number of parallel threads, and hence the number of partitions,
is determined at the beginning of each sweep 
by the percentage $\rho$ of suspended point insertions during the previous sweep.
If $\rho=1$, a single thread is used so that 
the termination of the algorithm is guaranteed.

\begin{figure}[hbt]
	\centering
	\begin{subfigure}[b]{0.2\textwidth}
		\centering
		\includegraphics[width=0.8\textwidth]{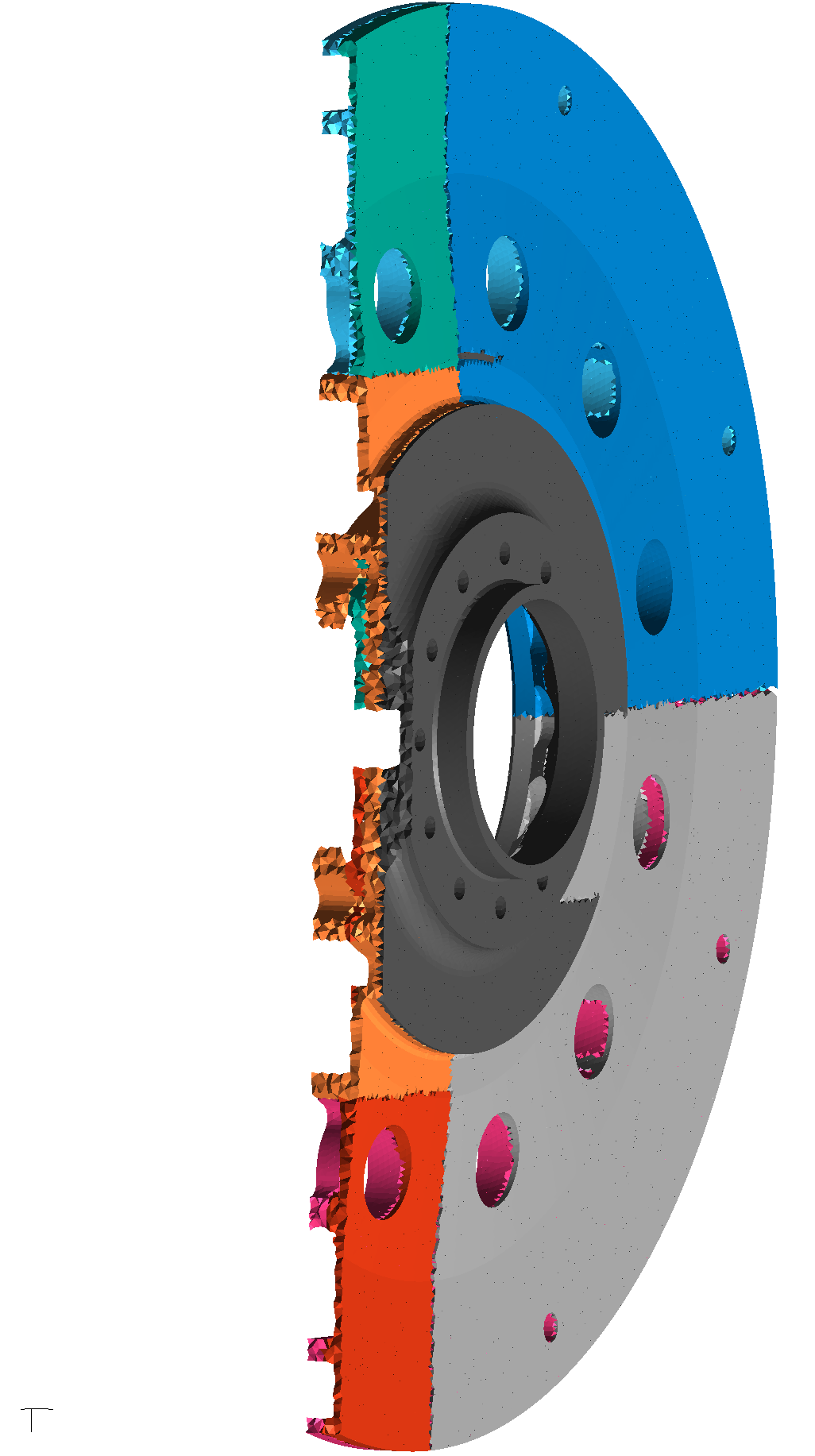}~~~
		\caption{Rotor mesh partitions}
		\label{fig:rotor_partitions}
	\end{subfigure}
	~~~
	\begin{subfigure}[b]{0.3\textwidth}
		\centering
		\includegraphics[width=0.8\textwidth]{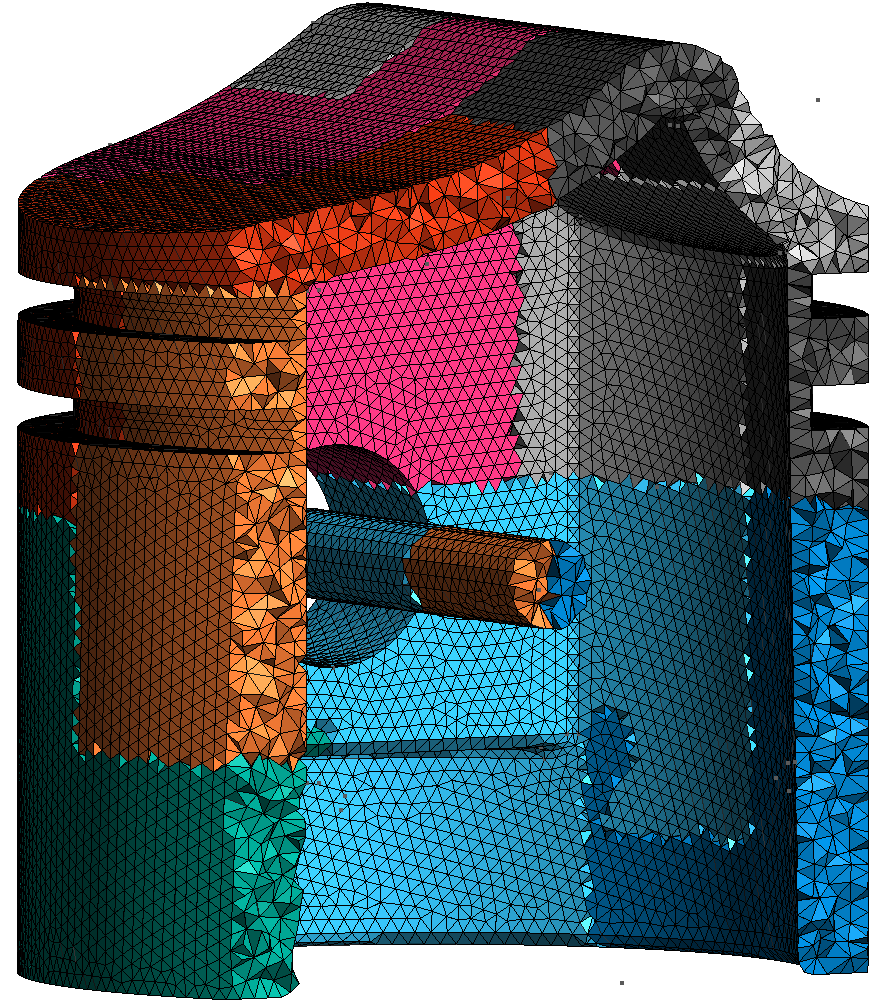}
		\caption{Piston mesh partitions}
	\end{subfigure}
	~~~
	\begin{subfigure}[b]{0.35\textwidth}
		\centering
		\includegraphics[width=0.8\textwidth]{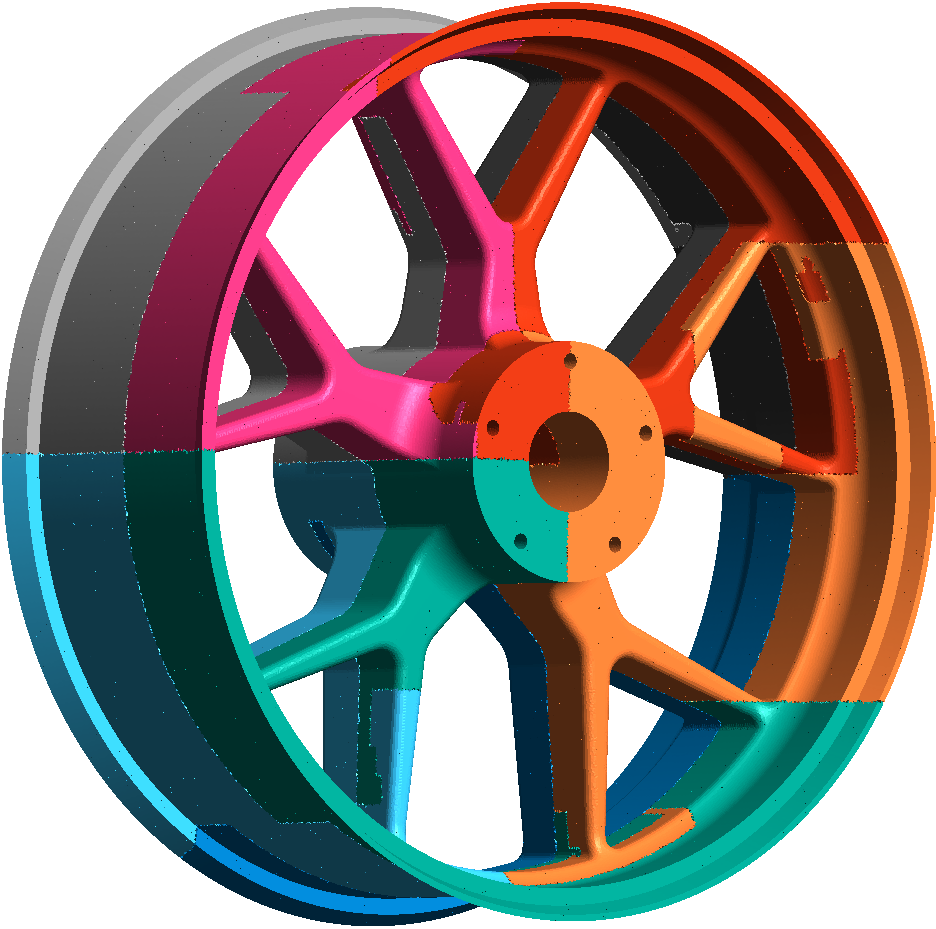}
		\caption{Rim mesh partitions}
	\end{subfigure}
    \caption{Partitions based on the 3D Moore curve ordering. These partitions
    were created almost instantaneously at the start of the mesh
    improvement step, running on 8 threads.}
	\label{fig:partitions}
\end{figure}

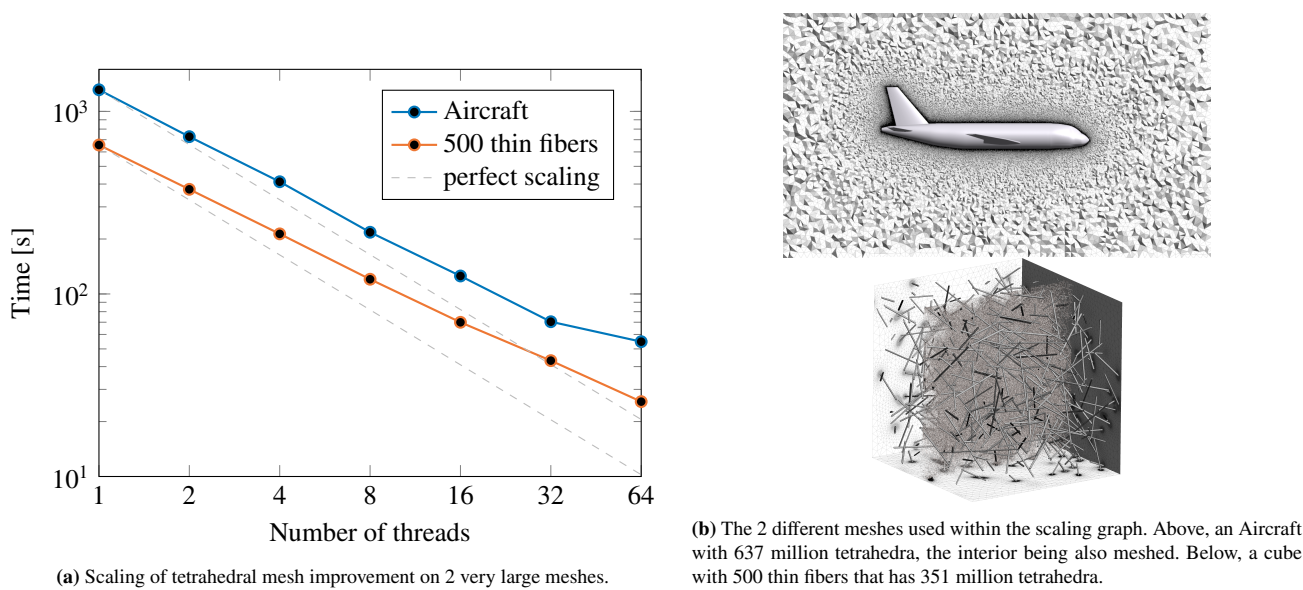
\begin{figure}[hbt]
\centering
\begin{subfigure}[b]{0.5\textwidth}
\centering
\begin{tikzpicture}
\begin{axis}[
scale only axis,
width=0.8\textwidth,
height=0.6\textwidth,
xmode=log,
xmin=1,
xmax=64,
cycle list={
  {draw=tolBlue},
  {draw=tolOrange}
},
xtick={1,2,4,8,16,32,64},
xticklabels={1,2,4,8,16,32,64},
xlabel={Number of threads},
ymode=log,
ymin=10,
ymax=1700,
ylabel={Time [s]},
legend style={at={(0.95,0.95)}, anchor=north east, legend cell align=left,align=left},
]

\addplot+[solid,thick, mark=*]
  table[row sep=crcr]{
1 1313.585 \\
2 727.2374000000001 \\
4 411.8236 \\
8 217.77579999999998 \\
16 125.4354 \\
32 70.4606 \\
64 54.782 \\
};
\addlegendentry{Aircraft};

\addplot+[solid,thick, mark=*]
  table[row sep=crcr]{
1 654.5484 \\
2 374.34540000000004 \\
4 213.37640000000002 \\
8 120.38319999999999 \\
16 69.9496 \\
32 43.130399999999995 \\
64 25.712 \\
};
\addlegendentry{500 thin fibers};


\addplot+[color=tolGrey,dashed, mark=none]
  table[row sep=crcr]{
1	1313.585 \\
64 20.524765625 \\
};
\addlegendentry{perfect scaling};

\addplot+[color=tolGrey,dashed, mark=none]
  table[row sep=crcr]{
1	654.5484 \\
64 10.22731875 \\
};


\end{axis}
\end{tikzpicture}
\caption{Scaling of tetrahedral mesh improvement on 2 very large meshes.}
\label{fig:scaling}
\end{subfigure}
~
\begin{subfigure}[b]{0.45\textwidth}
\centering
\includegraphics[height=0.4\textwidth]{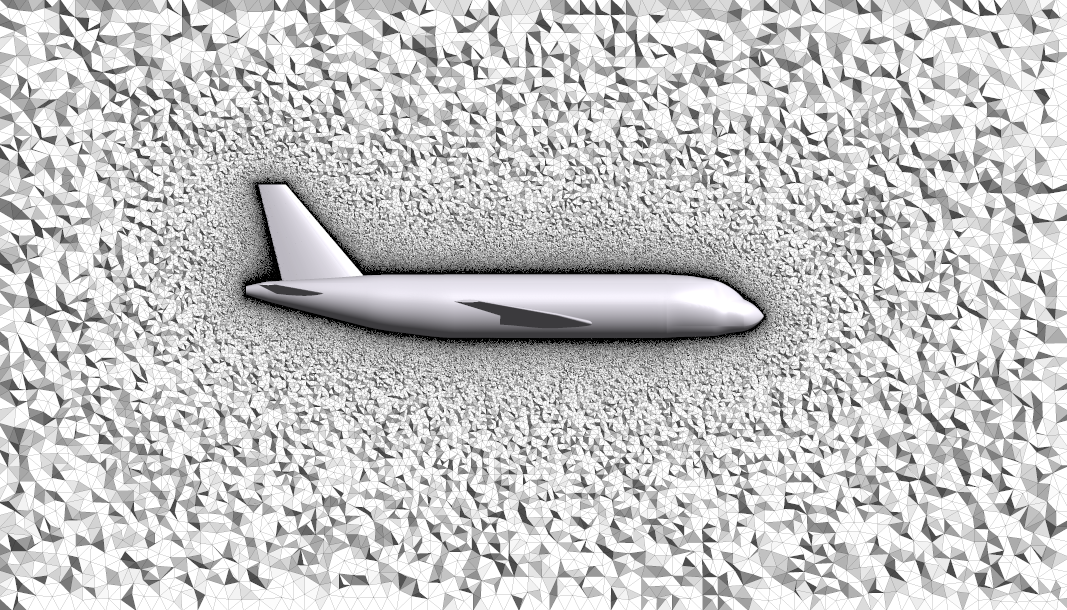}\\
\includegraphics[height=0.4\textwidth]{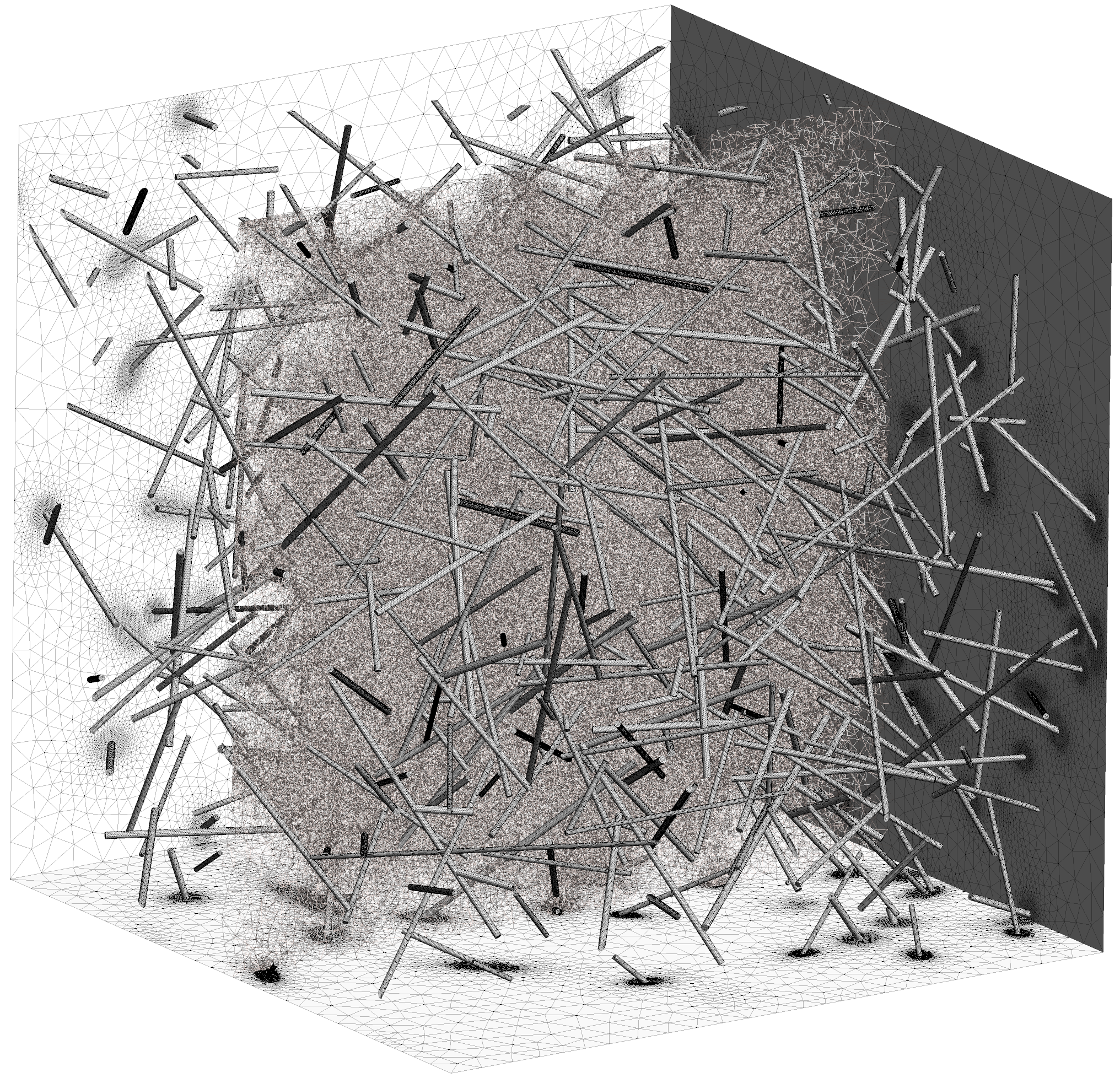}
\caption{The 2 different meshes used within the scaling graph. Above, an
Aircraft with 637 million tetrahedra, the interior being also meshed. Below, a
cube with 500 thin fibers that has 351 million tetrahedra.}
\label{fig:huge}
\end{subfigure}
\caption{Scaling of our parallel mesh improvement schedule on 2 huge meshes}
\end{figure}

Similarly, for mesh improvement, the space filling curve is partitioned so as to
equally distribute bad tetrahedra over the threads. Figure~\ref{fig:partitions}
shows partitioning examples for 8 threads on 3 different meshes. The number of
threads is again decided in function of the percentage of suspended
operations---due to a conflict with another partition--- in the previous sweep.
The SER loop is thus executed repeatedly with decreasing numbers of threads,
until all partition conflicts have been resolved. The same procedure is used for
the GSC loop as well. This parallel algorithm scales well in case of very big
meshes only, for essentially two reasons. Firstly, 
elements crossing partition boundaries 
represent a larger portion of space in small meshes than in large meshes,
yielding thus mechanically more conflicts. This is less of an issue with mesh
refinement because cavities are usually smaller for point insertion than for the
Growing SPR Cavity. The second reason is that the time spent in one execution of
the SER or of the GSC loop is very small, typically in the millisecond range.
Therefore, the overhead of launching new threads and computing Moore indices
for the whole mesh is significant. 
Figure~\ref{fig:scaling} compares the
scaling in the case of two highly-refined models. The mesh generation was done
on the 64 physical cores of an Intel Xeon Phi 7210 machine, running at 1.3 GHz.
The overall effectiveness and performance of our mesh improvement algorithm is
analyzed in section~\ref{subsec:improve}.

\section{Mesh generator's performance}\label{sec:4}

Before comparing the performance of our tetrahedral mesh generator with TetGen
and Gmsh, the similarities and key differences between
the different implementations are briefly recalled. 
Besides being all open-source and free, 
the three considered software tools are also structured very similarly. 
As explained in the introduction, the mesh generation process can be splitted 
into four distinct steps: creation of an \emph{empty mesh}, 
\emph{boundary recovery}, \emph{mesh refinement} and \emph{mesh improvement}.

Although it could be sensible to work on improving the quality of tetrahedra
right at the moment of their creation, none of the open source mesh generators
that are considered here use such a strategy. Separating mesh refinement and
mesh improvement has several advantages. The mesh refinement process essentially
creates around $4$ times the number of tetrahedra that are present in the final
mesh. Avoiding the computation of element quality during refinement therefore
allows a substantial gain in efficiency. Moreover, having separated pieces of
code for refinement and improvement allows to programmers to focus on smaller
tasks and goals. It also allows using meshing capabilities of our code in a
modular fashion: it is indeed possible to create a mesh with Gmsh, refine it
with TetGen, and optimize it with our HXT algorithm, without duplicating 
any part of the process.
Another advantage is that
the implementations of the different steps can be analyzed separately,
and their respective execution times be compared on different models.
The results of our benchmark for all 4 steps of the tetrahedral mesh generation, 
including the specification of relevant hardware characteristics and compilation flags,
are given in Appendix~\ref{bench}. More specifically, element numbers, timings
and CPU/memory usages are reported in Table~\ref{tab:models}. Comparative
performances per step, except for the boundary recovery step, and per mesh generator
are shown as bar plots in Figure~\ref{fig:barplot}.

\subsection{Empty mesh}

The first step, i.e., the creation of the empty mesh, consists in computing
a Delaunay
tetrahedralization through all points of the surface mesh plus some possible
user-defined interiors points. Since a Delaunay tetrahedralization is unique,
provided a \emph{simulation of
simplicity}\cite{DBLP:journals/tog/EdelsbrunnerM90} is used in conjunction with
robust adaptive predicates\cite{shewchuk1997adaptive}, all three programs
generate identical empty meshes. The ordering of the tetrahedra
however, can largely vary from one packege to another. 
The ordering is even non-deterministic with HXT, because
threads can reserve chunks of memory 
corresponding to a set of 8192 tetrahedra
at different moments from one run to another. 
To get rid of this non-deterministic behaviour, 
HXT offers a \emph{reproducible} mode, which reorders deterministically
tetrahedra in the lexicographic order of their nodes. 
The performance of TetGen, Gmsh and HXT
(with and without the reproducible mode),
for the different meshes shown in Table~\ref{tab:models},
are reported in Figure~\ref{fig:emptymesh}. 
HXT is the fastest for the empty mesh step, but not because of 
parallelization. Most tetrahedra of the empty mesh indeed
cross the domain from side to side. With our partitioning method based on
a space-filling curve, points that are at different extremities of the domain
have very little chances of being in the same partition. Because a tetrahedron
is only considered in a partition when at least three of its vertices are in that
partition, parallelization is not very effective at this step. The speed
difference between HXT and TetGen, or between HXT and Gmsh, 
is rather explained here by the good serial performances of our Delaunay
kernel\cite{marot_one_nodate}.


\subsection{Boundary Recovery}

All 3 software tools 
rely internally on TetGen's boundary recovery code. 
Table~\ref{tab:models} shows however that HXT is a bit slower
than TetGen for boundary recovery, 
because of the back and forth conversion of the mesh
between its own data structure and TetGen's format. 
Gmsh applies the same,
albeit much slower, kind of conversion. 
Moreover, the order under which tetrahedra are stored in memory 
can either slow down or speedup the boundary
recovery process, which explains timing differences
observed between HXT in normal or \emph{reproducible} mode. 
As the other parts of HXT are parallelized and optimized for heavy workload, 
boundary recovery is the bottleneck of our code 
for large meshes. 
In contrast, HXT usually performs very well on small meshes. We suspect that
TetGen's algorithm for locating missing facets or edges has a superlinear
complexity with respect to the size of the mesh.

\subsection{Mesh Refinement}

Mesh refinement is the step where our parallel HXT algorithm really stands out. 
The ways TetGen, HXT or Gmsh proceed to refine the mesh are rather different, 
but Figure~\ref{fig:numtet} nonetheless shows that the different software tools 
generate meshes with only slightly different numbers of tetrahedra. 
TetGen was given the options \textcmd{q1.1/14}, 
causing it to refine only tetrahedra with a radius-edge ratio larger than
$1.1$ or a minimum dihedral angle smaller than
$14^{\circ}$\cite{noauthor_tetgen_nodate}. 
No meshsize constraint was given, although TetGen allows it. 
TetGen therefore adds point only to optimize the quality of elements. 
In contrast, HXT and Gmsh add a new point $p_{k+1}$ 
inside a tetrahedron $\tau$ if the insertion 
does not create an edge shorter than the prescribed meshsize, 
which is 
the value linearly interpolated
from meshsizes at the vertices of $\tau$. 
At the beginning of the refinement step, 
nodal meshsizes 
are evaluated from the surface mesh.
and Gmsh thus do not refine with the goal of improving the quality of the elements. 
All three softwares do however end up with meshes with very similar numbers of tetrahedra.
Still, timings for mesh refinement and improvement indicated in Table~\ref{tab:models},
Figure~\ref{fig:refine} and \ref{fig:optimize}, are expressed in second per million
tetrahedra to alleviate the effect of the different refinement strategies.
HXT is at least 5 times faster than Gmsh and TetGen for mesh
refinement, consistently 
generating more than one million tetrahedra per second
when the other mesh generators peak at $300\,000$ tetrahedra per second. On the
\texttt{300 fibers} model, HXT reaches nearly 3 million tetrahedra per
second, which is 132 times faster than Gmsh. 
HXT is also efficient in terms of memory usage.  
For the \texttt{100 fibers} mesh, the peak resident set size 
does not exceed 77 bytes per tetrahedron. 
If memory allocations not depending on the size of the mesh are put aside,
HXT consumes only about 60 bytes per tetrahedron.

\subsection{Mesh Improvement}\label{subsec:improve}

For the mesh improvement step, TetGen was given the \textcmd{-o/130} option,
setting  to $130^\circ$ the target maximum \emph{dihedral angle}. In contrast,
HXT and Gmsh target a minimum \emph{inradius to longest edge ratio}
$\gamma_{threshold}=0.35$. As Gmsh and HXT avoid modifying the surface mesh, 
TetGen was also prevented from modifying the surface mesh by enabling the
\textcmd{-Y} option. In addition to the timings of Table~\ref{tab:models},
mesh quality has also been compared after the mesh improvement step 
for 3 geometries (\emph{Rotor}, \emph{Piston} and \emph{Rim}) 
and 3 quality measures:

\begin{itemize}
	\item the dihedral angles, plotted in Figure~\ref{fig:hist-dihedral}
	\item the inradius to longest edge ratio ($\gamma$), shown on the bar plot
     of Figure~\ref{fig:hist-gamma}. The tetrahedra with $\gamma < 0.35$ before
     and after HXT's mesh improvement step are also 
     shown in Figure~\ref{fig:gamma035}
	\item the signed inverse condition number (SICN), 
      plotted in Figure~\ref{fig:hist-sicn}
\end{itemize}

HXT gives comparatively the best results regarding each of those three quality measures, 
even for the maximum dihedral angles which is 
the quality measure used  by TetGen during the mesh improvements step. 
Figure~\ref{fig:hist-dihedral} indeed shows
noticeably smaller maximum dihedral angles for HXT compared to Gmsh and TetGen
for each of the 3 considered meshes. This difference is explained by the
effectiveness of the new \emph{Growing SPR Cavity} operation described in
section \ref{sec:2}. The performance of HXT
is even more striking when looking at the minimum inradius to
longest edge ratio and the minimum SICN:

\begin{table}[!h]
	\centering
	\begin{tabular}{l c c c c c c c c c c c c c c c}
	 & \multicolumn{4}{c}{Rotor} & & \multicolumn{4}{c}{Piston} & & \multicolumn{4}{c}{Rim}\\
	 & $\gamma$ & SICN & $\measuredangle_{min}$ & $\measuredangle_{max}$ & & $\gamma$ & SICN & $\measuredangle_{min}$ & $\measuredangle_{max}$ & & $\gamma$ & SICN & $\measuredangle_{min}$ & $\measuredangle_{max}$ \\\toprule
	Gmsh & 0.014 & 0.077 & 4.06 & 174.76 & & 0.23 & 0.23 & 11.24 & 161.40 & & 0.0094 & 0.034 & 1.59 & 177.22\\
	TetGen & 0.037 & 0.16 & 5.10 & 168.28 & & 0.15 & 0.30 & 14.40 & 156.12 & & 0.073 & 0.15 & 6.70 & 168.26\\
	HXT (ours) & 0.13 & 0.16 & 5.10 & 149.2 & & 0.29 & 0.39 & 13.76 & 146.64 & & 0.16 & 0.17 & 5.92 & 149.09\\
	\end{tabular}
	\caption{Minimum $\gamma$ and SICN measure among all tetrahedra, minimum and maximum dihedral angle among all dihedral angles of the mesh, for the
	\emph{Rotor}, \emph{Piston} and \emph{Rim} meshes with the 3 different open-source
	software tools tested.}
\end{table}

HXT's mesh improvement is parallelized and has very fast smoothing and edge
removal operations. 
Table~\ref{tab:models} and Figure \ref{fig:optimize} indeed indicate
that HXT is also the fastest 
when it comes to mesh improvement, but not by much. 
The reason for this is simple:
TetGen and Gmsh stop optimizing whenever the smoothing and edge removal operations become ineffective,  
whereas HXT, then, starts its first GSC sweep.

\section{Conclusion and discussion}

This paper has presented the parallel tetrahedral mesh generator HXT, 
and demonstrated its efficiency by means of a detailed comparative benchmark
with two concurrent open-source software tools: Gmsh and TetGen. 
Although the obtained performance are very satisfactory,
there is room for improvement. 
The first and easiest improvement
will be the implementation of edge contraction, point insertion and 2-2 flips. 
Another particularly challenging task 
is the parallelization of the boundary recovery step,
which is the last non-parallelized step in the whole meshing process.
Thanks to the acquired experience,
improvement on the boundary recovery procedures of TetGen 
is now definitely within reach,
probably but adding our parallelization strategy based on Moore curves. 
The novel GSC operation could also help gaining performance,
and could be adapted as a last resort operation to recover constrained
facets or edges. 
Other parallelizations strategies should also be tested for the mesh improvement
stage.

\section*{Acknowledgments}

This project has received funding from the European Research Council (ERC) under
the European Union's Horizon 2020 research and innovation programme (grant
agreement ERC-2015-AdG-694020). We are grateful to the authors of the ABC
Dataset\cite{Koch_2019_CVPR}, for providing the \emph{Rim} and \emph{Piston}
geometries (respectively model 40 and 9733). Surface meshes were generated
thanks to Gmsh.

\bibliography{biblio}

\newpage
\appendix
\section{Benchmarks}
\label{bench}
Benchmarks use the average of 5 runs on an Intel$^\circledR$ Core\texttrademark
i7-6700HQ with 4 cores running at 3.5GHz and 16Gb of RAM. We use Gmsh 4.6.0 with
the \texttt{-optimize\_threshold=0.35} option and HXT's \texttt{tetMesh\_CLI}
executable provided in
\href{https://gitlab.onelab.info/gmsh/gmsh/-/tree/master/contrib/hxt/tetMesh/}{gmsh/contrib/hxt/tetMesh/}
with default options. We use TetGen 1.5.1 with \texttt{"-BNEFIVVYp -q1.1/14 -O7
-o/130"} options. TetGen was compiled with \texttt{gcc -O3}, Gmsh and HXT with
\texttt{gcc -O3 -march=native}. See section \ref{sec:4} for an in-depth
discussion on the results. The code for the benchmarks is available at
\url{https://git.immc.ucl.ac.be/marotc/tetmesher_benchmark}

\subsection{Speed}\label{bench:speed}
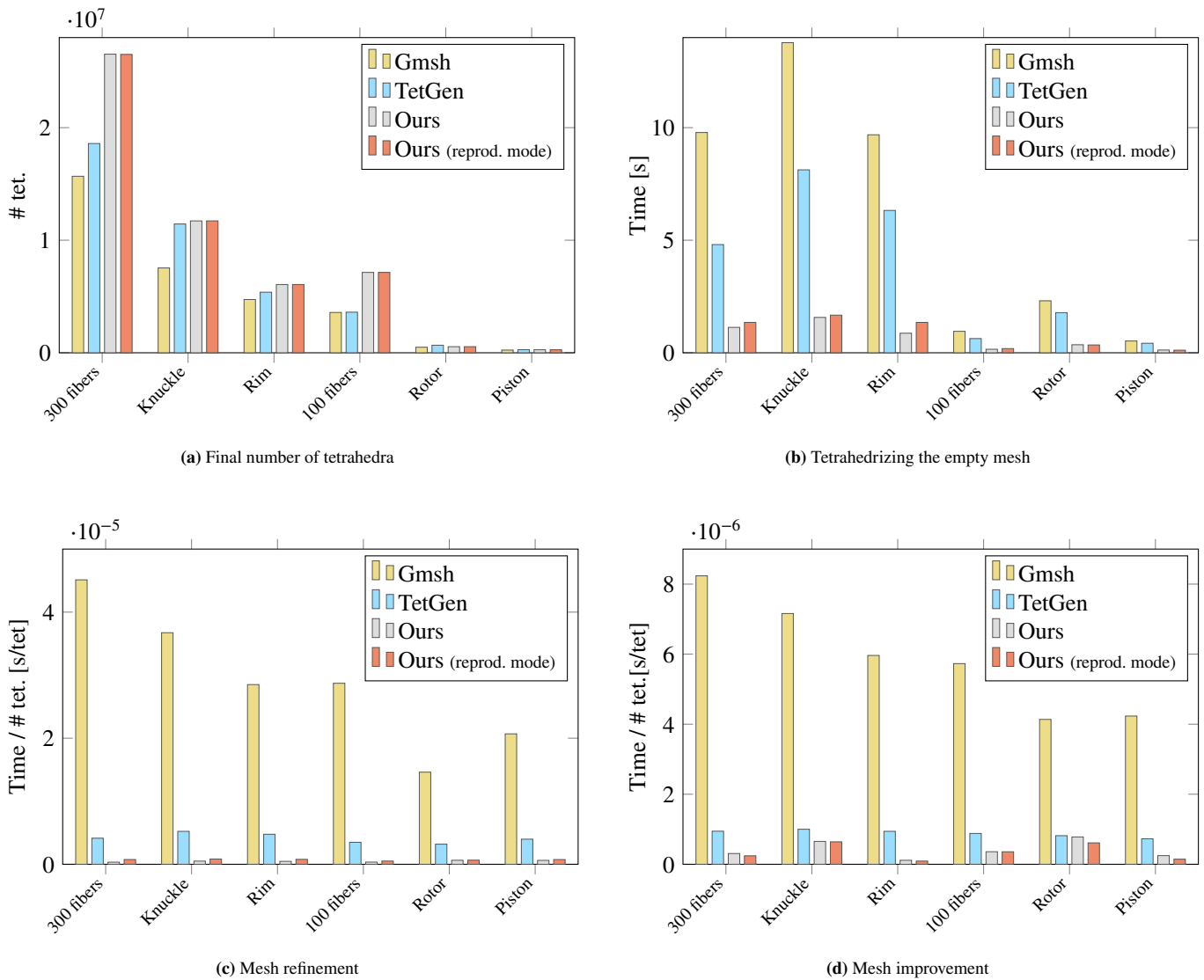
\begin{figure}[!hb]
	\centering
\begin{subfigure}[b]{0.48\textwidth}
\centering
\begin{tikzpicture}
\begin{axis}[%
  scale only axis,
  width=0.9\textwidth,
  height=0.55\textwidth,
  ybar,
  ymax=28000000,
  ymin=0,
  bar width=0.02\textwidth, 
  cycle list={
    {fill=tolLightYellow, line width=0mm, draw=tolDarkGrey},
    {fill=tolLightCyan, line width=0mm, draw=tolDarkGrey},
    {fill=tolPaleGrey, line width=0mm, draw=tolDarkGrey},
    {fill=tolLightOrange, line width=0mm, draw=tolDarkGrey}
  },
  ylabel={\# tet.},
  symbolic x coords={%
    fibers300,
    fuseecurv,
    rim,
    fibers100,
    rotor,
    piston,
  },
  xtick={
    fibers300,
    fuseecurv,
    rim,
    fibers100,
    rotor,
    piston,
  },
  xticklabels={%
    300 fibers,
    Knuckle,
    Rim,
    100 fibers,
    Rotor,
    Piston,
  },
  x tick label style={rotate=45,anchor=east,align=center,font=\footnotesize},
  y label style={at={(axis description cs:-0.05,.5)},anchor=south},
  table/col sep=comma,
  legend cell align={left},
  filter discard warning=false, 
]

\addplot
  table[x=file,y=numTets, discard if not={command}{Gmsh}]{data/speed.csv}; 
\addlegendentry{Gmsh};

\addplot
  table[x=file,y=numTets, discard if not={command}{TetGen}]{data/speed.csv}; 
\addlegendentry{TetGen};

\addplot
  table[x=file,y=numTets, discard if not={command}{Hxt}]{data/speed.csv}; 
\addlegendentry{Ours};

\addplot
  table[x=file,y=numTets, discard if not={command}{HxtRep}]{data/speed.csv}; 
\addlegendentry{Ours {\footnotesize(reprod. mode)}};
\end{axis}
\end{tikzpicture}
\caption{Final number of tetrahedra}
\label{fig:numtet}
\end{subfigure}
\hspace{0.03\textwidth}
\begin{subfigure}[b]{0.48\textwidth}
\centering
\begin{tikzpicture}
\begin{axis}[%
  scale only axis,
  width=0.9\textwidth,
  height=0.55\textwidth,
  ybar,
  ymax=14,
  ymin=0,
  bar width=0.02\textwidth, 
  cycle list={
    {fill=tolLightYellow, line width=0mm, draw=tolDarkGrey},
    {fill=tolLightCyan, line width=0mm, draw=tolDarkGrey},
    {fill=tolPaleGrey, line width=0mm, draw=tolDarkGrey},
    {fill=tolLightOrange, line width=0mm, draw=tolDarkGrey}
  },
  ylabel={Time [s]},
  symbolic x coords={%
    fibers300,
    fuseecurv,
    rim,
    fibers100,
    rotor,
    piston,
  },
  xtick={
    fibers300,
    fuseecurv,
    rim,
    fibers100,
    rotor,
    piston,
  },
  xticklabels={%
    300 fibers,
    Knuckle,
    Rim,
    100 fibers,
    Rotor,
    Piston,
  },
  x tick label style={rotate=45,anchor=east,align=center,font=\footnotesize},
  y label style={at={(axis description cs:-0.05,.5)},anchor=south},
  table/col sep=comma,
  legend cell align={left},
  filter discard warning=false, 
]

\addplot
  table[x=file,y=emptyMesh, discard if not={command}{Gmsh}]{data/speed.csv}; 
\addlegendentry{Gmsh};

\addplot
  table[x=file,y=emptyMesh, discard if not={command}{TetGen}]{data/speed.csv}; 
\addlegendentry{TetGen};

\addplot
  table[x=file,y=emptyMesh, discard if not={command}{Hxt}]{data/speed.csv}; 
\addlegendentry{Ours};

\addplot
  table[x=file,y=emptyMesh, discard if not={command}{HxtRep}]{data/speed.csv}; 
\addlegendentry{Ours {\footnotesize(reprod. mode)}};
\end{axis}
\end{tikzpicture}
\caption{Tetrahedrizing the empty mesh}
\label{fig:emptymesh}
\end{subfigure}
\\[0.75cm]
\begin{subfigure}[b]{0.48\textwidth}
\centering
\begin{tikzpicture}
\begin{axis}[%
  scale only axis,
  width=0.9\textwidth,
  height=0.55\textwidth,
  ybar,
  ymax=0.00005,
  ymin=0,
  bar width=0.02\textwidth,
  cycle list={
  {fill=tolLightYellow, line width=0mm, draw=tolDarkGrey},
  {fill=tolLightCyan, line width=0mm, draw=tolDarkGrey},
  {fill=tolPaleGrey, line width=0mm, draw=tolDarkGrey},
  {fill=tolLightOrange, line width=0mm, draw=tolDarkGrey}
  },
  ylabel={Time / \# tet. [s/tet]},
  symbolic x coords={%
    fibers300,
    fuseecurv,
    rim,
    fibers100,
    rotor,
    piston,
  },
  xtick={
    fibers300,
    fuseecurv,
    rim,
    fibers100,
    rotor,
    piston,
  },
  xticklabels={%
    300 fibers,
    Knuckle,
    Rim,
    100 fibers,
    Rotor,
    Piston,
  },
  x tick label style={rotate=45,anchor=east,align=center,font=\footnotesize},
  y label style={at={(axis description cs:-0.05,.5)},anchor=south},
  table/col sep=comma,
  legend cell align={left},
  filter discard warning=false, 
]

\addplot
  table[x=file,y expr={\thisrow{refine}/\thisrow{numTets}}, discard if not={command}{Gmsh}]{data/speed.csv}; 
\addlegendentry{Gmsh};

\addplot
  table[x=file,y expr={\thisrow{refine}/\thisrow{numTets}}, discard if not={command}{TetGen}]{data/speed.csv}; 
\addlegendentry{TetGen};

\addplot
  table[x=file,y expr={\thisrow{refine}/\thisrow{numTets}}, discard if not={command}{Hxt}]{data/speed.csv}; 
\addlegendentry{Ours};

\addplot
  table[x=file,y expr={\thisrow{refine}/\thisrow{numTets}}, discard if not={command}{HxtRep}]{data/speed.csv}; 
\addlegendentry{Ours {\footnotesize(reprod. mode)}};
\end{axis}
\end{tikzpicture}
\caption{Mesh refinement}
\label{fig:refine}
\end{subfigure}
\hspace{0.03\textwidth}
\begin{subfigure}[b]{0.48\textwidth}
\centering
\begin{tikzpicture}
\begin{axis}[%
  scale only axis,
  width=0.9\textwidth,
  height=0.55\textwidth,
  ybar,
  ymax=0.000009,
  ymin=0,
  bar width=0.02\textwidth, 
  cycle list={
    {fill=tolLightYellow, line width=0mm, draw=tolDarkGrey},
    {fill=tolLightCyan, line width=0mm, draw=tolDarkGrey},
    {fill=tolPaleGrey, line width=0mm, draw=tolDarkGrey},
    {fill=tolLightOrange, line width=0mm, draw=tolDarkGrey}
  },
  ylabel={Time / \# tet.[s/tet]},
  symbolic x coords={%
    fibers300,
    fuseecurv,
    rim,
    fibers100,
    rotor,
    piston,
  },
  xtick={
    fibers300,
    fuseecurv,
    rim,
    fibers100,
    rotor,
    piston,
  },
  xticklabels={%
    300 fibers,
    Knuckle,
    Rim,
    100 fibers,
    Rotor,
    Piston,
  },
  x tick label style={rotate=45,anchor=east,align=center,font=\footnotesize},
  y label style={at={(axis description cs:-0.05,.5)},anchor=south},
  table/col sep=comma,
  legend cell align={left},
  filter discard warning=false, 
]

\addplot
  table[x=file,y expr={\thisrow{opti}/\thisrow{numTets}}, discard if not={command}{Gmsh}]{data/speed.csv}; 
\addlegendentry{Gmsh};

\addplot
  table[x=file,y expr={\thisrow{opti}/\thisrow{numTets}}, discard if not={command}{TetGen}]{data/speed.csv}; 
\addlegendentry{TetGen};

\addplot
  table[x=file,y expr={\thisrow{opti}/\thisrow{numTets}}, discard if not={command}{Hxt}]{data/speed.csv}; 
\addlegendentry{Ours};

\addplot
  table[x=file,y expr={\thisrow{opti}/\thisrow{numTets}}, discard if not={command}{HxtRep}]{data/speed.csv}; 
\addlegendentry{Ours {\footnotesize(reprod. mode)}};
\end{axis}
\end{tikzpicture}
\caption{Mesh improvement}
\label{fig:optimize}
\end{subfigure}
	\caption{Performance benchmark bar plots}
	\label{fig:barplot}
\end{figure}

Figure \ref{fig:barplot} shows bar plots of different performance measures,
further detailed in Table \ref{tab:models}.

\begin{table}[!hb]
\begin{tabular}{r c c c c c c}
\multirow{2}{*}{models\vspace{\fill}}
&
\includegraphics[height=2.3cm]{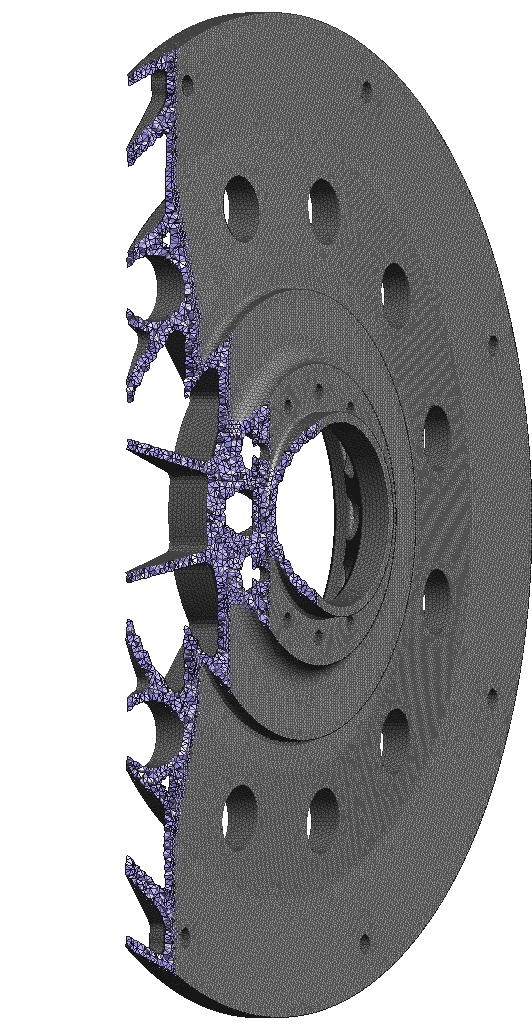} &
\includegraphics[height=2cm]{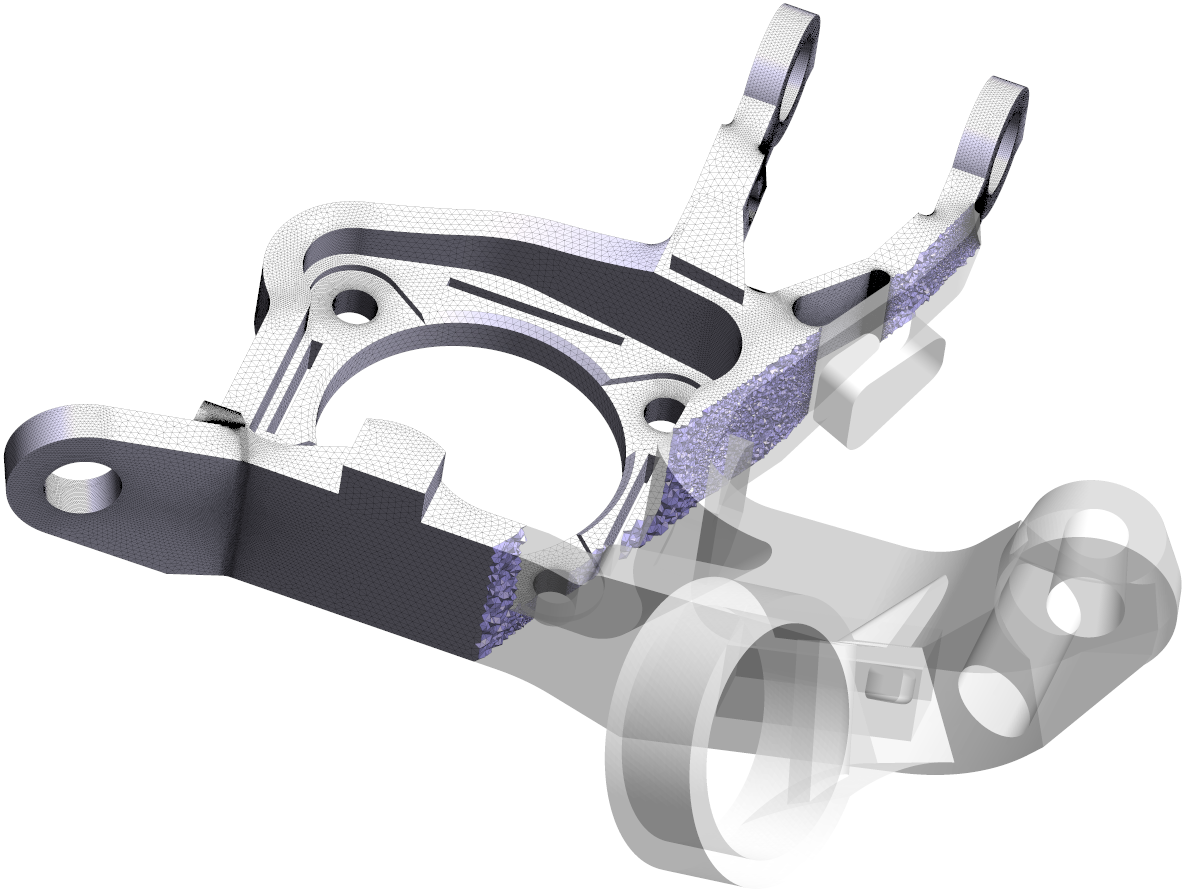} &
\includegraphics[height=2.3cm]{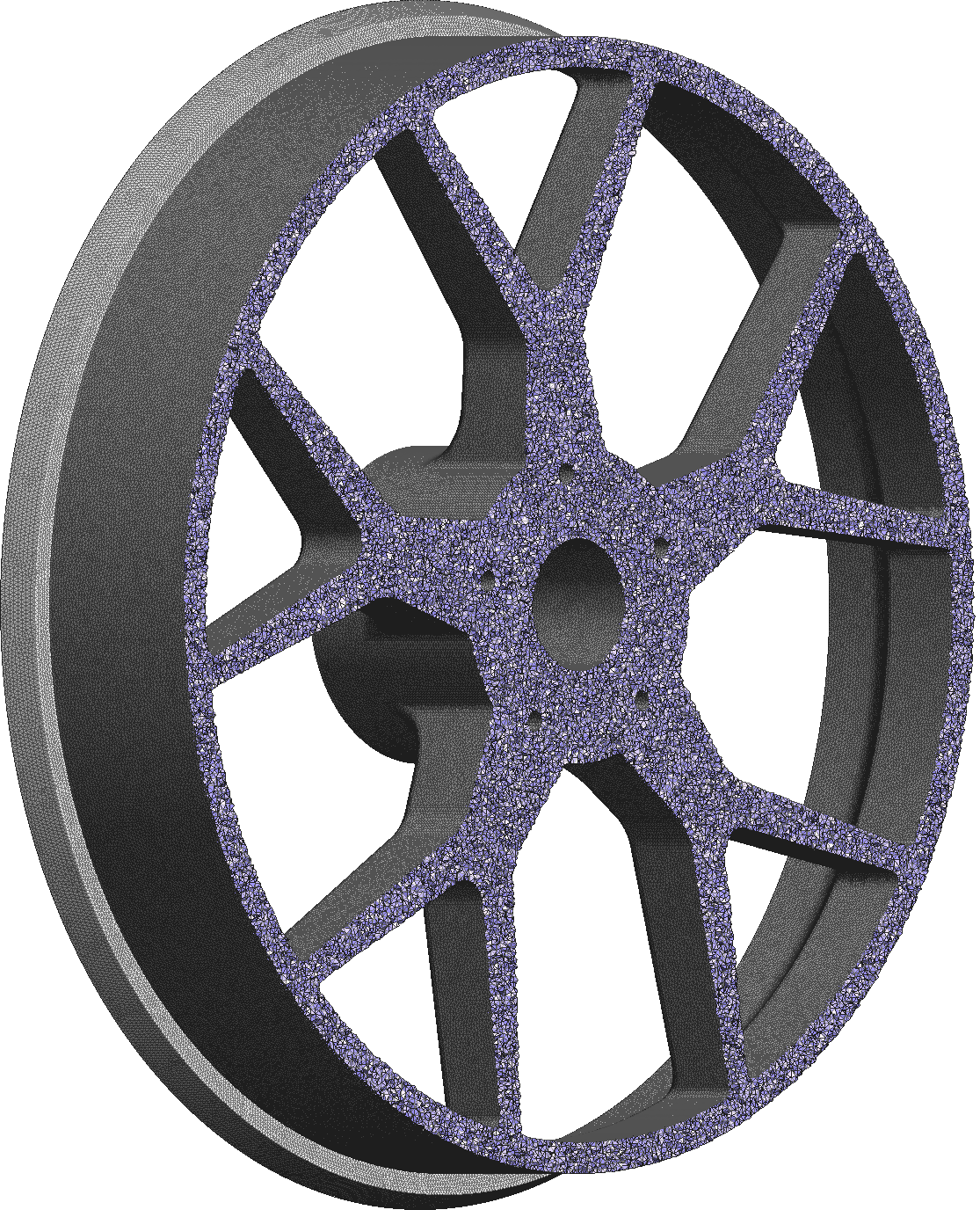} &
\multicolumn{2}{c}{\includegraphics[height=2.3cm]{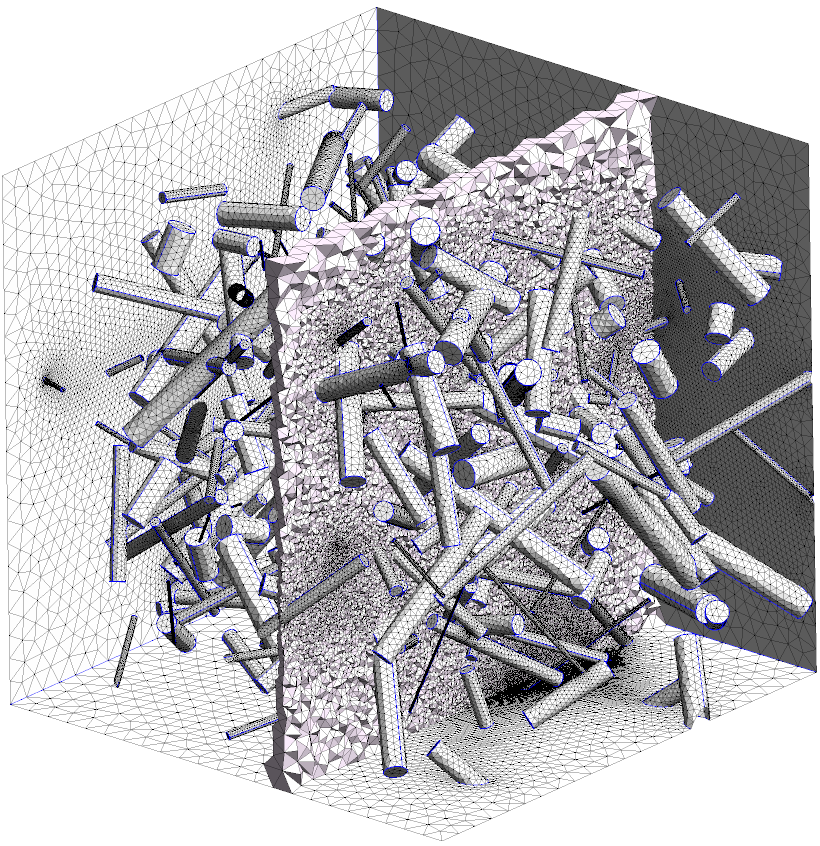}} &
\includegraphics[height=2.3cm]{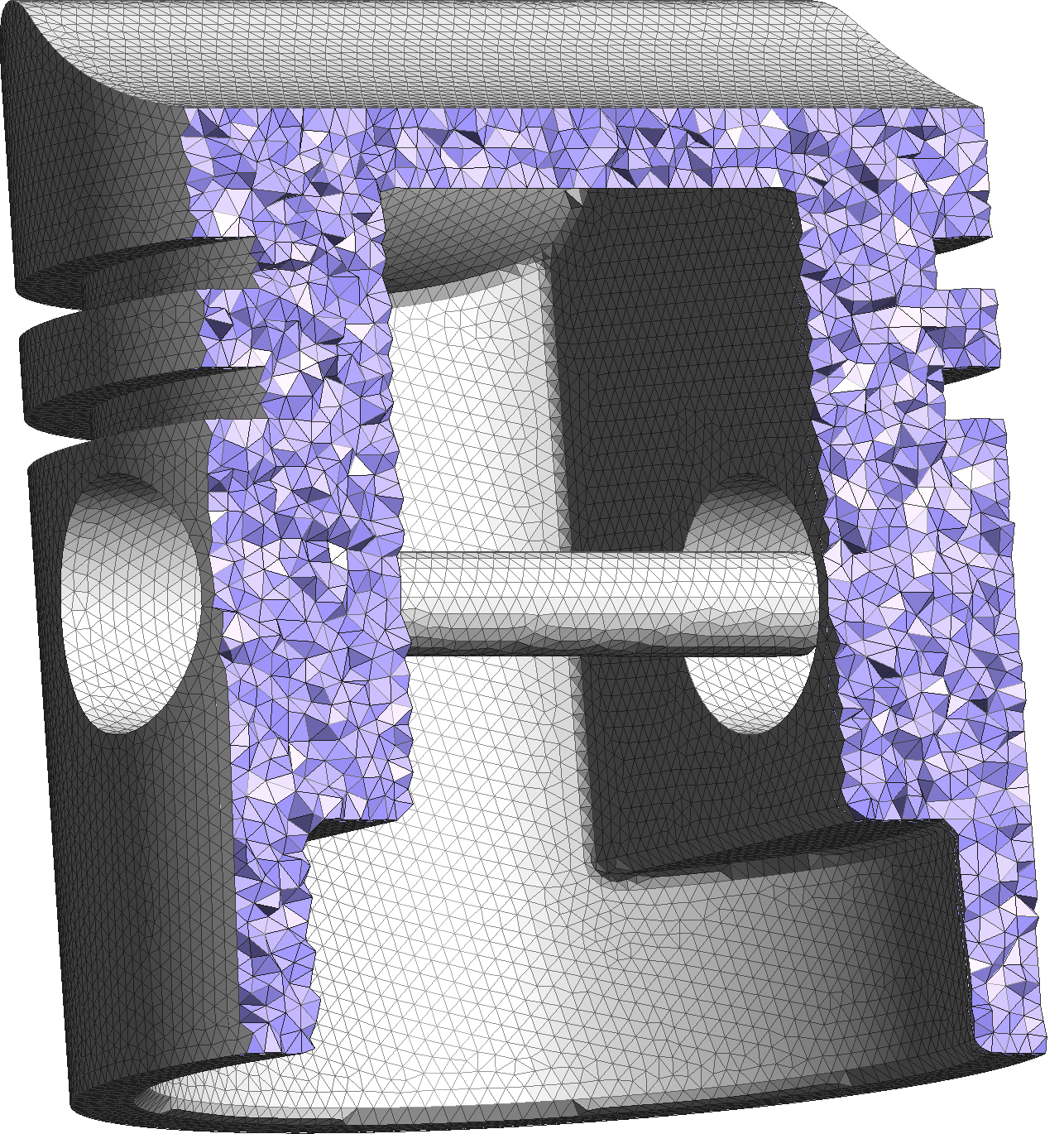} \\[0.1cm]
 & Rotor & Knuckle & Rim & 100 fibers & 300 fibers & Piston\\\toprule
\textbf{Surface mesh} & & & & & & \\
number of points & $115\,052$ & $435\,423$ & $397\,985$ & $47\,759$ & $328\,661$ & $27\,187$\\
number of triangles & $230\,232$ & $870\,914$ & $796\,030$ & $94\,994$ & $656\,162$ & $54\,374$\\[0.3cm]
\textbf{Gmsh} & & & & & & \\
final number of points & $138\,652$ & $1\,435\,736$ & $964\,213$ & $583\,860$ & $2\,583\,840$ & $52\,823$\\
final number of tetrahedra & $493\,632$ & $7\,536\,855$ & $4\,729\,987$ & $3\,580\,914$ & $15\,680\,789$ & $241\,982$\\
Empty Mesh [$s$] & 2.31 & 13.78 & 9.68 & 0.95 & 9.78 & 0.52\\
Boundary Recovery [$s$] & 2.43 & 33.039 & 32.93 & 6.30 & 0.07 & 23.92\\
Mesh Refinement [$\mu s/\text{tet}$] & 14.61 & 36.73 & 28.50 & 28.72 & 45.12 & 20.69\\
Mesh Improvement [$\mu s/\text{tet}$] & 4.14 & 7.16 & 5.96 & 5.73 & 8.24 & 4.23\\
Max. mem. usage [$B/\text{tet}$]  & 1226.70 & 558.86 & 569.25 & 524.24 & 493.53 & 865.92\\
CPU \% & 99.9 & 99.9 & 99.9 & 99.9 & 99.9 & 99.9\\[0.15cm]
\textbf{TetGen} & & & & & & \\
final number of points & $169\,755$ & $2\,140\,715$ & $1\,101\,619$ & $614\,938$ & $3\,183\,439$ & $59\,876$ \\
final number of tetrahedra & $670\,594$ & $11\,442\,634$ & $5\,382\,975$ & $3\,606\,324$ & $18\,589\,149$ & $275\,491$ \\
Empty Mesh [$s$] & 1.78 & 8.13 & 6.32 & 0.63 & 4.81 & 0.42 \\
Boundary Recovery [$s$] & 1.23 & 6.21 & 4.70 & 0.54 & 4.60 & 0.32 \\
Mesh Refinement [$\mu s/\text{tet}$] & 3.19 & 5.23 & 4.76 & 3.49 & 4.14 & 3.98 \\
Mesh Improvement [$\mu s/\text{tet}$] & 0.82 & 1.00 & 0.94 & 0.88 & 0.94 & 0.73 \\
Max. mem. usage [$B/\text{tet}$]  & 903.46 & 314.12 & 474.32 & 204.72 & 227.18 & 614.19 \\
CPU \% & 99.9 & 99.9 & 99.9 & 99.9 & 99.9 & 99.9\\[0.15cm]
\textbf{HXT (ours)} & & & & & & \\
final number of points & $146\,957$ & $2\,121\,555.4$ & $1\,184\,497.6$ & $1\,162\,217$ & $4\,350\,880.2$ & $57\,452$ \\
final number of tetrahedra & $541\,929.6$ & $11\,714\,033.8$ & $6\,063\,127.6$ & $7\,139\,662$ & $26\,526\,744.2$ & $268\,116.4$ \\
Empty Mesh [$s$] & 0.36 & 1.57 & 0.87 & 0.16 & 1.13 & 0.12\\
Boundary Recovery [$s$] & 1.91 & 8.54 & 6.64 & 0.72 & 6.53 & 0.47\\
Mesh Refinement [$\mu s/\text{tet}$] & 0.64 & 0.50 & 0.47 & 0.35 & 0.34 & 0.62\\
Mesh Improvement [$\mu s/\text{tet}$] & 0.78 & 0.65 & 0.11 & 0.36 & 0.31 & 0.25\\
Max. mem. usage [$B/\text{tet}$]  & 1032.73 & 221.15 & 328.88 & 80.86 & 77.13 & 646.72\\
CPU \% & 292.2 & 383.0 & 326.4 & 523.8 & 479.4 & 351.2\\[0.3cm]
\textbf{HXT (ours)} & & & & & & \\
\textbf{reproducible mode} & & & & & & \\
final number of points & $146\,967$ & $2\,121\,944$ & $1\,184\,569$ & $1\,162\,848$ & $4\,348\,050$ & $57\,505$ \\
final number of tetrahedra & $542\,012$ & $11\,716\,723$ & $6\,064\,318$ & $7\,141\,786$ & $26\,506\,921$ & $268\,493$ \\
Empty Mesh [$s$] & 0.35 & 1.67 & 1.35 & 0.19 & 1.35 & 0.12\\
Boundary Recovery [$s$] & 1.60 & 7.60 & 6.10 & 0.64 & 5.74 & 0.42\\
Mesh Refinement [$\mu s/\text{tet}$] & 0.65 & 0.83 & 0.78 & 0.51 & 0.75 & 0.75\\
Mesh Improvement [$\mu s/\text{tet}$] & 0.61 & 0.64 & 0.90 & 0.36 & 0.24 & 0.15\\
Max. mem. usage [$B/\text{tet}$] & 1031.42 & 221.40 & 328.60 & 123.45 & 113.98 & 654.47\\
CPU \% & 301.4 & 444.6 & 391.4 & 577.4 & 514.8 & 379.2\\
\end{tabular}
\caption{Performance benchmark table}
\label{tab:models}
\end{table}

\clearpage
\subsection{Dihedral Angles}

The dihedral angles formed by each pair of facets of a tetrahedron are customary
measures to look at because of their conceptual simplicity. However, this
measure does not directly correspond to any type of error that the
discretization induces during a finite element simulation \cite{shewchuk_what_2002}.

\begin{figure}[!hb]
	\input{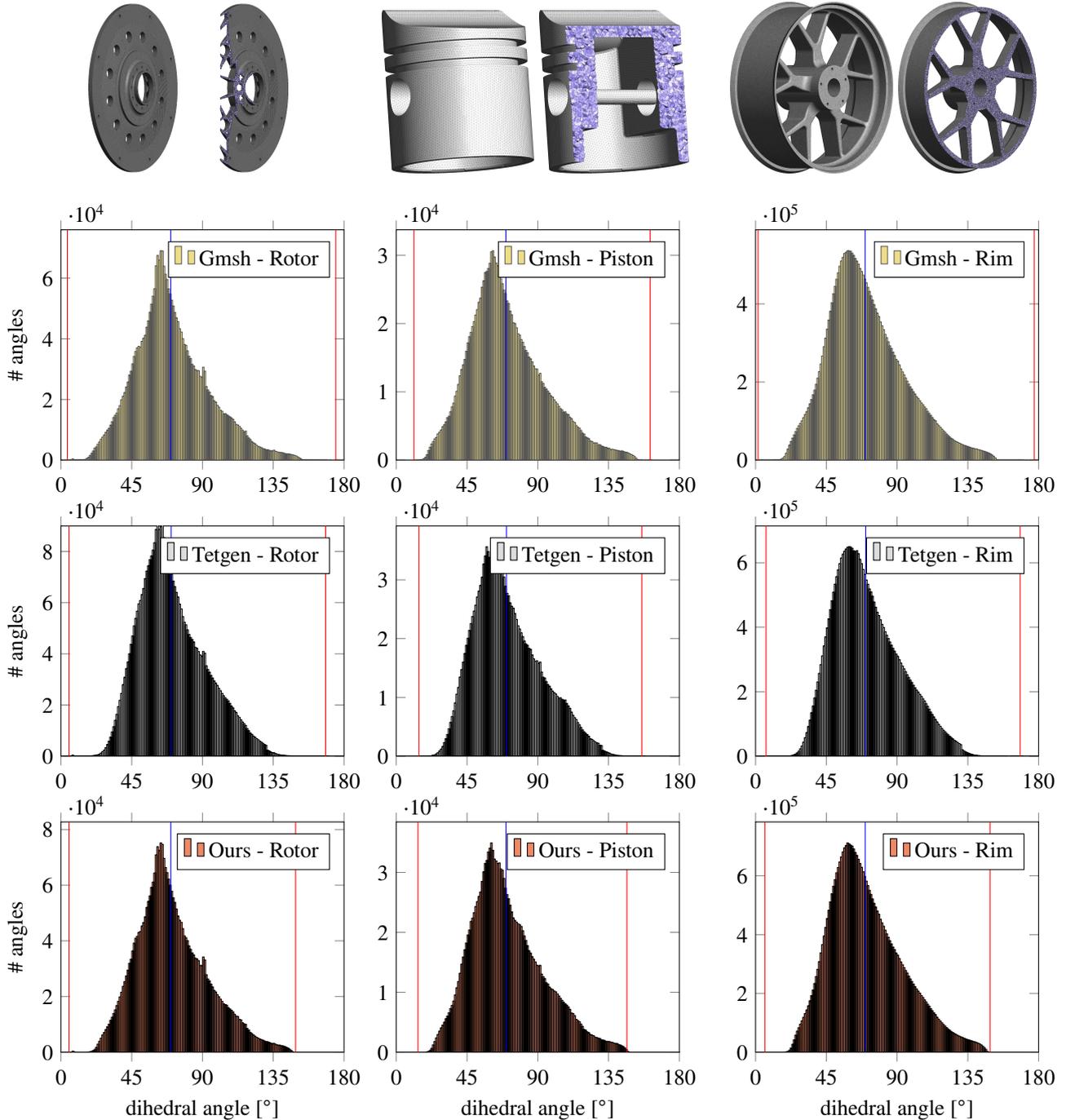}
	\caption{Histograms of the dihedral angles. Lower and upper bound are
	marked with red vertical lines. The average is marked with a blue line.}
	\label{fig:hist-dihedral}
\end{figure}

\subsection{Gamma}
\label{bench:gamma}

\begin{equation*}
	\gamma = \frac{\sqrt{24}~3V}{|e_{max}|(A_1 + A_2 + A_3 + A_4)}
	       = \frac{\sqrt{24}~r_{in}}{|e_{max}|},
\end{equation*}
where $V$ is the volume of the tetrahedron, $|e_{max}|$ is the length of the
longest edge, $A_i$ is the area of the $i$th face and $r_{in}$ is the inradius
of the tetrahedron.  The factor $\sqrt{24}$ is added such that the optimal
tetrahedron, which is a regular tetrahedron, has a quality $\gamma = 1$. This
quality measure penalize all tetrahedra according to their associated
interpolation error \cite{shewchuk_what_2002}.

\begin{figure}[!hb]
	\input{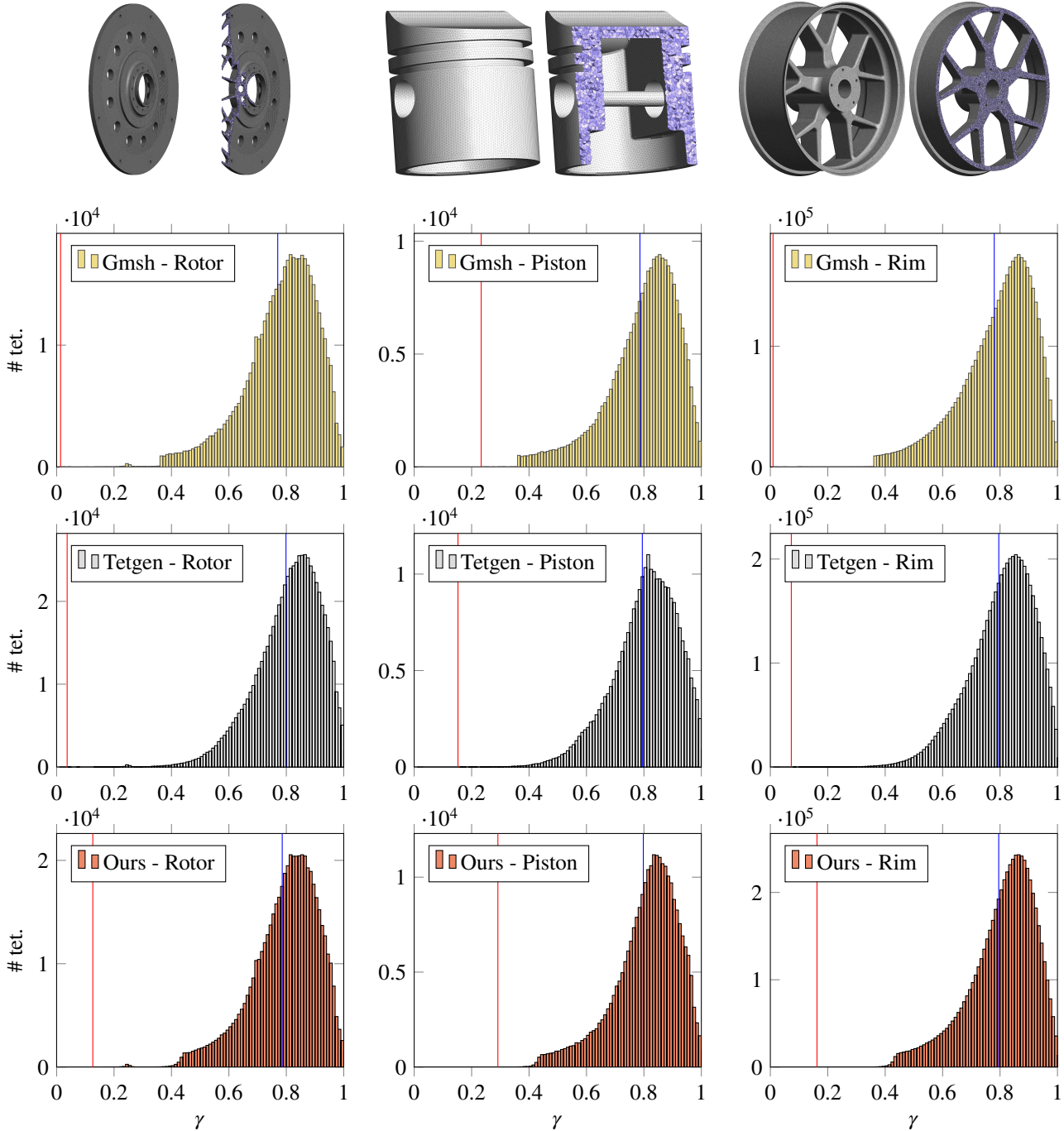}
	\caption{Histograms of the quality measure $\gamma$. Lower bound is marked
	with a red vertical line. The average is marked with a blue line.}
	\label{fig:hist-gamma}
\end{figure}

\subsection{Signed inverse condition number (SICN)}
\label{bench:sicn}

The inverse condition number in Frobenius norm
$\text{SICN}=\frac{3}{\kappa(S)}$ for each tetrahedron. $\kappa(S)$ is the
condition number of the linear transformation matrix $S$ between a tetrahedron
of the mesh and a regular tetrahedron. the SICN is directly proportional to the
greatest lower bound for the distance of $S$ to the set of singular
matrices \cite{freitag_tetrahedral_2002}.

\begin{figure}[!hb]
	\input{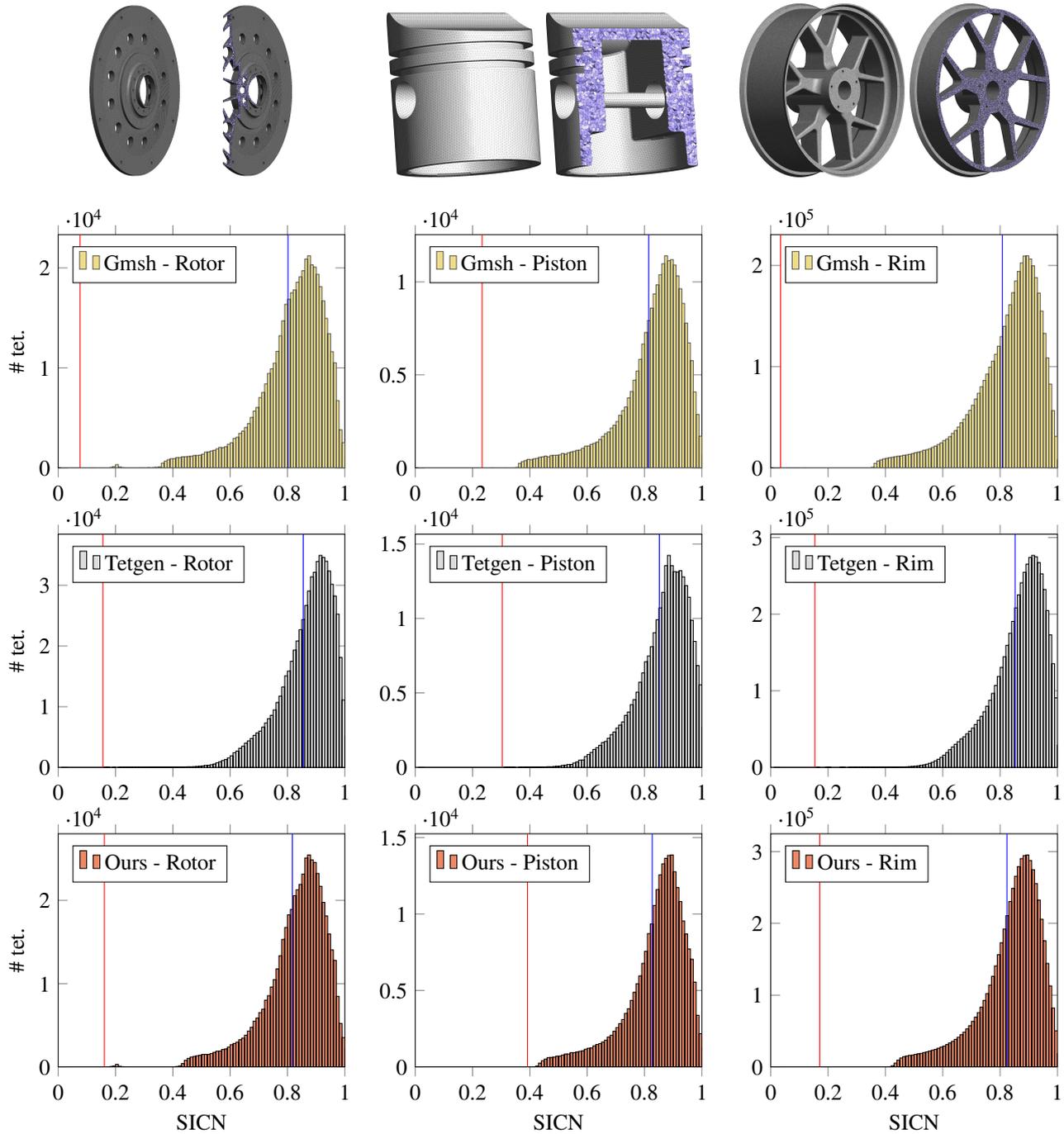}
	\caption{Histograms of the SICN quality measure. Lower bound is marked
	with a red vertical lines. The average is marked with a blue line.}
	\label{fig:hist-sicn}
\end{figure}




\end{document}